\documentstyle[12pt,epsf]{article}

\advance \topmargin by -\headheight
\advance \topmargin by -\headsep

\evensidemargin \oddsidemargin
\def\ie{{\em i.e.}}

\def\ie{\hbox{\it i.e.}}

\def\CC{{\mathchoice
{\rm C\mkern-8mu\vrule height1.45ex depth-.05ex 
width.05em\mkern9mu\kern-.05em}
{\rm C\mkern-8mu\vrule height1.45ex depth-.05ex 
width.05em\mkern9mu\kern-.05em}
{\rm C\mkern-8mu\vrule height1ex depth-.07ex 
width.035em\mkern9mu\kern-.035em}
{\rm C\mkern-8mu\vrule height.65ex depth-.1ex 
width.025em\mkern8mu\kern-.025em}}}

\def\RR{{\rm I\kern-1.6pt {\rm R}}}

\def\ZZ{{\rm Z}\kern-3.8pt {\rm Z} \kern2pt}

\def\np{Nucl. Phys.}
\def\pl{Phys. Lett.}
\def\prl{Phys. Rev. Lett.}
\def\pr{Phys. Rev.}

\def\ijmp{Int. J. Mod. Phys.}
\def\mpl{Mod. Phys. Lett.}

\def\atmp{Adv. Theor. Math. Phys. }
\def\jhep{J. High Energy Phys.}

\def\jgp{J. Geom. Phys.}
\def\atmp{Adv. Theor. Math. Phys.}

\newcommand{\beq}{\begin{equation}}
\newcommand{\eeq}{\end{equation}}
\newcommand{\rc}{\nonumber\\}
\newcommand{\bear}{\begin{eqnarray}}
\newcommand{\eear}{\end{eqnarray}}

\newfont{\namefont}{cmr10}
\newfont{\addfont}{cmti7 scaled 1440}
\newfont{\boldmathfont}{cmbx10}
\newfont{\headfontb}{cmbx10 scaled 1728}
\renewcommand{\theequation}{{\rm\thesection.\arabic{equation}}}
\begin{document}
\begin{titlepage}

\begin{center} \Large \bf Stable Wrapped Branes

\end{center}

\vskip 0.3truein
\begin{center} 
J. M. Camino
\footnote{e-mail:camino@fpaxp1.usc.es}, 
A. Paredes
\footnote{e-mail:angel@fpaxp1.usc.es}
and 
A.V. Ramallo
\footnote{e-mail:alfonso@fpaxp1.usc.es}

\vspace{0.3in}

Departamento de F\'\i sica de
Part\'\i culas, \\ Universidad de Santiago\\
E-15706 Santiago de Compostela, Spain. 
\vspace{0.3in}

\end{center}
\vskip 1truein

\begin{center}
\bf ABSTRACT
\end{center} 

We study some wrapped configurations of branes in the near-horizon
geometry of a stack of other branes. The common feature of all  the cases 
analyzed is a quantization rule and the  appearance of a finite number of
static configurations in which the branes  are partially wrapped on
spheres. The energy of these configurations can be given in closed form
and the analysis of their small oscillations shows that they are stable.
The cases studied include D(8-p)-branes in the type II supergravity
background of Dp-branes for $0\le p\le 5$, M5-branes in the M5-brane
geometry in M-theory and D3-branes in a $(p,q)$ fivebrane background in
the type IIB theory.  The brane configurations found admit the
interpretation of bound states of strings (or M2-branes in M-theory)
which extend along the unwrapped directions. We check this fact directly
in a particular case by using the Myers polarization mechanism.

\vskip4.5truecm
\leftline{US-FT-3/01 \hfill April 2001}
\leftline{hep-th/0104082}
\smallskip
\end{titlepage}
\setcounter{footnote}{0}


\setcounter{equation}{0}
\section{Introduction}
\medskip

In a recent paper \cite{Bachas},  Bachas, Douglas and Schweigert have
shown how D-branes on a group manifold  are stabilized against shrinking
(see also ref. \cite{Pavel}). The concrete model studied in ref.
\cite{Bachas} was the motion of a D2-brane in the geometry of the $SU(2)$
group manifold. Topologically,
$SU(2)$ is equivalent to a three-sphere $S^3$. The D2-brane is embedded
in this $S^3$ along a two-sphere $S^2$ which, in a system of spherical
coordinates, is placed at constant latitude angle $\theta$.
The D2-brane dynamics is determined by the Born-Infeld action, in which a
worldvolume gauge field is switched on. An essential ingredient in the
analysis of ref. \cite{Bachas} is the quantization condition of the
worldvolume flux, which, with our notations, can be written as:
\beq
\int_{S^2}\,F\,=\,{2\pi n\over T_f}\,\,,
\,\,\,\,\,\,\,\,\,\,\,\,\,\,\,\,\,\,
n\in\ZZ\,\,,
\label{uno}
\eeq
where $F$ is the worldvolume gauge field strength and $T_{f}$ is the
tension of the fundamental string, which, in terms of the Regge slope
$\alpha'$ is  $T_{f}\,=\,( 2\pi\alpha\,'\,)^{-1}\,$. 

By using eq. (\ref{uno}) one can easily find the form of the 
worldvolume gauge field strength for non-zero $n$, and the corresponding
value of the energy of the D2-brane. The minimum of this energy
determines the embedding of the brane in the group manifold, which
occurs at a finite set of latitude angles $\theta$. It turns out that the
static configurations found by this method are stable under small
perturbations and exactly match
those obtained by considering open strings on group  manifolds
\cite{KS, KO}. In this
latter approach the D-brane configurations are determined by all the
possible boundary conditions of the corresponding Conformal Field Theory
(CFT). Actually \cite{Ale, FFFS}, 
each possible boundary condition corresponds to a
D-brane wrapped on a (twisted) conjugacy class of the group.

The underlying CFT imposes quantization conditions on the allowed  
(twisted) conjugacy classes, which can be interpreted geometrically in
terms of the embedding of the D-brane worldvolume in the group manifold.
Thus, for example, in the case of the $SU(2)$ group manifold, the
non-trivial conjugacy classes are two-spheres embedded in 
$SU(2)\approx S^3$. The quantization conditions of the corresponding
Wess-Zumino-Witten (WZW) model determine that only a finite number of 
$S^2\subset S^3$ embeddings are possible and  their number is
related to the level of the affine $su(2)$ Kac-Moody algebra \cite{Ale,
FFFS}. Actually, to each conjugacy class one associates a  Cardy boundary
state \cite{Cardy} of the WZW model. The mass of the D-brane
configuration can be obtained, in this approach, by computing \`a la
Polchinski \cite{Polchi} the matrix element between Cardy states of the
string theory cylinder diagram, and comparing the result with  the one
obtained in a gravitational field theory. Apart from a finite shift in
the level of the current algebra, the mass obtained in this way is
exactly the same as the one computed with the Born-Infeld action and the
quantization condition (\ref{uno}). For other aspects of this open string
approach and of the flux quantization condition (\ref{uno}) see
ref. \cite{all}.

The agreement between the Born-Infeld and CFT approaches for the system
of ref. \cite{Bachas} is quite remarkable. For this reason the
generalization of this result to other backgrounds and brane probes is
very interesting. The $SU(2)$ group manifold studied in \cite{Bachas} can
be regarded as a component of the transverse part of a Neveu-Schwarz (NS)
fivebrane geometry. Thus, the natural generalization to consider is a
Ramond-Ramond (RR) background. This case was studied in ref.
\cite{PR}, where it was shown that the brane probe must be partially
wrapped on some angular directions and extended along the radial
coordinate. 

Following the analysis of ref. \cite{PR}, in this paper we  study,
first of all, the motion of a D(8-p)-brane in the background of a stack
of parallel Dp-branes. The external region of the Dp-brane metric has
$SO(9-p)$ rotational symmetry, which is manifest when a system of
spherical coordinates is chosen. In this system of coordinates a
transverse $S^{8-p}$ sphere is naturally defined and the constant latitude
condition on the $S^{8-p}$ determines a $S^{7-p}$ sphere. We shall embed
the D(8-p)-brane in this background in such a way that it is wrapped on
this $\,S^{7-p}\subset S^{8-p}$ constant latitude sphere and extended
along the radial direction. Therefore, as in ref. \cite{Bachas},
the brane configuration is characterized by an angle $\theta$, which
parametrizes  the latitude of the  $S^{7-p}$. 

In order to analyze this Dp-D(8-p) system by means of the Born-Infeld
action, we shall establish first some quantization condition which,
contrary to (\ref{uno}),  will now involve the electric components of the
worldvolume gauge field. By using this quantization rule we shall find a
finite set of stable brane configurations characterized by some angles
$\theta$ which generalize the ones found in ref. \cite{Bachas}. The
energy of these configurations will be also computed and, from this
result, we shall conclude that semiclassically our D(8-p)-brane
configurations can be regarded as a bound state of fundamental strings.
On the other hand, we will find a first order BPS differential equation
whose fulfillment implies the saturation of an energy bound and whose
constant $\theta$ solutions are precisely our wrapped configurations.
This BPS equation is the one \cite{Imamura}  satisfied by the baryon
vertex \cite{Wittenbaryon}, which will allow us to interpret our
configurations as a kind of short distance limit (in the radial
direction) of the baryonic branes \cite{CGS,Craps,Camino}.

Another purpose of this paper is to study a mechanism of flux
stabilization in M-theory. We shall consider, in particular, a M5-brane
probe in a M5-brane background. By using the Pasti-Sorokin-Tonin (PST)
\cite{PST} action for the M5-brane probe, we shall look for static
configurations in which the probe is wrapped on a three-sphere. After
establishing a flux quantization condition similar to (\ref{uno}), we
shall find these configurations and we will show that they closely
resemble  those found for the D4-D4 system. Actually, our states can be
interpreted semiclassically as BPS bound states of M2-branes and they are
related to the short distance limit of the baryonic vertex of
M-theory \cite{Ali,kappa}. 

Another example which we will work out in detail is the one in which the
background is a stack of fivebranes which have both NS and RR charges,
\ie\ a collection of the so-called $(p,q)$ fivebranes \cite{LuRoy}. In
this case the probe is a D3-brane and the ``magnetic" quantization
condition (\ref{uno}) and our electric generalization must be imposed at
the same time. We will show that these two quantization rules are indeed
compatible and we will find the stable wrapped configurations of the
D3-brane probe. Again, they can be interpreted semiclassically as a
collection of strings (actually, in this case,  $(q,p)$ strings). 

If our brane configurations admit an interpretation as bound states of
strings (or M2-branes in the case of M-theory), it should be possible to
obtain them starting directly from the strings (or M2-branes). We will
check this fact in a particular case. Indeed, we will show
how one can build up the wrapped D3-brane configurations in the 
NS fivebrane background by using D-strings
in the same background. The mechanism
responsible for this transmutation is the one advocated by
Myers \cite{Myers}, in which the D-strings move in a noncommutative
fuzzy sphere and are polarized by the background. 

This paper is organized as follows. In section 2 we will study the 
Dp-D(8-p) system. Section 3 is devoted to the analysis of the flux
stabilization in M-theory. The D3-brane in the $(p,q)$ fivebrane
background is considered in section 4. In section 5 we summarize our
results and explore some directions for future work. The paper is
completed with two appendices. In appendix A we collect the functions
which determine the location of the wrapped brane configurations in the
transverse sphere. In appendix B we show how to obtain the wrapped
D3-branes from polarized D-strings.

\setcounter{equation}{0}
\section{Wrapped branes in Ramond-Ramond backgrounds}
\medskip

The ten-dimensional metric corresponding to a stack of $N$ coincident
extremal Dp-branes in the near-horizon region is given by
\cite{supergravity}:
\beq
ds^2\,=\,\Bigl[\,{r\over R}\,\Bigr]^{{7-p\over 2}}\,\,
(\,-dt^2\,+\,dx_{\parallel}^2\,)\,+\,
\Bigl[\,{R\over r}\,\Bigr]^{{7-p\over 2}}\,\,
(\,dr^2\,+\,r^2\,d\Omega_{8-p}^2\,)\,\,,
\label{dos}
\eeq
where $x_{\parallel}$ represent $p$ cartesian
coordinates along the branes, $r$ is a radial coordinate
parametrizing the distance to the branes and $d\Omega_{8-p}^2$ is the
line element of an unit $8-p$ sphere. We have written the metric in
the string frame. The parameter $R$, which we will refer to as 
the radius,   is given by:
\beq
R^{7-p}\,=\,N\,g_s\,2^{5-p}\,\pi^{{5-p\over 2}}\,
(\,\alpha\,'\,)^{{7-p\over 2}}\,\,
\Gamma\Bigl(\,{7-p\over 2}\Bigr)\,\,,
\label{tres}
\eeq
where $N$ is the number of Dp-branes of the stack and $g_s$ is the
string coupling constant. The metric (\ref{dos}) is a classical
solution of the type II supergravity equations of motion. This
solution is
also characterized by some non-vanishing values of the dilaton field
$\phi(r)$ and of a  Ramond-Ramond (RR) (8-p)-form field strength
$F^{(8-p)}$, namely:
\bear
e^{-\tilde\phi(r)}\,&=&\,\Bigl[\,{R\over r}\,
\Bigr]^{{(7-p)(p-3)\over 4}}
\,\,,\rc\rc
F^{(8-p)}\,&=&\,(7-p)\,R^{7-p}\,\epsilon_{(8-p)}\,\,,
\label{cuatro}
\eear
where $\tilde\phi(r)\,=\,\phi(r)\,-\,\phi(r\rightarrow {\infty})$, and we
are representing the Dp-brane as a magnetically charged object under the
$F^{(8-p)}$ form. In eq. (\ref{cuatro})(and in what follows) 
$\epsilon_{(n)}$ denotes the volume form of the sphere $S^n$. 

Let $\theta^1$, $\theta^2$, $\cdots$, $\theta^{8-p}$ be coordinates
which parametrize the $S^{8-p}$ transverse sphere. 
We shall assume that the $\theta$'s are spherical angles on $S^{8-p}$
and that $\theta\equiv\theta^{8-p}$ is the polar angle 
($0\le\theta\le\pi$). Therefore, the $S^{8-p}$ line element 
$d\Omega_{8-p}^2$ can be decomposed as:
\beq
d\Omega_{8-p}^2\,=\,d\theta^2\,+\,(\,{\rm sin}\,\theta)^{2}\,\,
d\Omega_{7-p}^2\,\,.
\label{cinco}
\eeq
In these coordinates it is not difficult to find a potential for the
RR gauge field. Indeed, let us  define the function $C_p(\theta)$ 
as the solution of the differential
equation:
\beq
{d\over d\theta}\, C_p(\theta)\,=\,-(7-p)\,({\rm sin}\,\theta)^{7-p}\,\,,
\label{seis}
\eeq
with the initial condition
\beq
C_p(0)\,=\,0\,\,.
\label{siete}
\eeq
It is clear that one can find by elementary integration a unique
solution to the problem of eqs. (\ref{seis}) and (\ref{siete}). Thus 
$C_p(\theta)$ can be considered as a known function of the polar angle
$\theta$. In terms of $C_p(\theta)$,   the RR potential $C^{(7-p)}$
can be represented as:
\beq
C^{(7-p)}\,=\,-R^{(7-p)}\,C_p(\theta)\,\,\epsilon_{(7-p)}\,\,.
\label{ocho}
\eeq
By using eq. (\ref{seis}) it can be easily verified that
\footnote{For simplicity, through this paper we choose the orientation
of the transverse $S^{8-p}$  sphere such that 
$\epsilon_{(8-p)}=(\sin\theta)^{7-p}\,d\theta\wedge\epsilon_{(7-p)}$.}: 
\beq
F^{(8-p)}\,=\,d\,C^{(7-p)}\,\,.
\label{nueve}
\eeq
Let us now consider a D(8-p)-brane embedded along the transverse
directions of the stack of  Dp-branes. The 
action of such a brane probe is  the sum of a
Dirac-Born-Infeld and a Wess-Zumino term:
\beq
S\,=\,-T_{8-p}\,\int d^{\,9-p}\sigma\,e^{-\tilde\phi}\,
\sqrt{-{\rm det}\,(\,g\,+\,F\,)}\,+\,
T_{8-p}\,\int \,\, F\wedge\,C^{(7-p)}\,\,,
\label{diez}
\eeq
where $g$ is the induced metric on the worldvolume of the D(8-p)-brane
and $F$ is a worldvolume abelian gauge field strength. The coefficient 
$T_{8-p}$ in eq. (\ref{diez}) is the tension of the D(8-p)-brane,
given by:
\beq
T_{8-p}\,=\,(2\pi)^{p-8}\,(\,\alpha\,'\,)^{{p-9\over 2}}\,
(\,g_s\,)^{-1}\,\,.
\label{once}
\eeq
The worldvolume
coordinates $\sigma^{\alpha}$ ($\alpha\,=\,0\,,\,\cdots\,,\,8-p\,)$ will
be taken as:
\beq
\sigma^{\alpha}\,=\,(\,t\,,\,r\,,\,
\theta^1\,,\,\cdots\,,\,\theta^{7-p}\,\,)\,\,.
\label{doce}
\eeq
\begin{figure}
\centerline{\hskip -.8in \epsffile{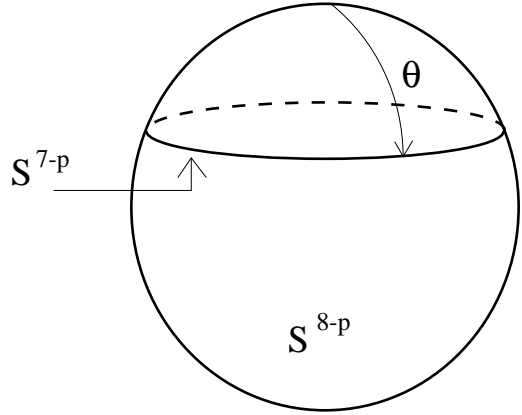}}
\caption{The points of the  $S^{8-p}$ sphere with the same polar angle
 $\theta$ define a  $S^{7-p}$ sphere. The angle $\theta$ represents the
latitude on $S^{8-p}$,  measured from one of its poles.
}
\label{fig1}
\end{figure}

With this election the embedding of the brane probe is described by a
function $\theta=\theta(\sigma^{\alpha})$. Notice that the
hypersurface $\theta\,=\,{\rm constant}$ defines a $S^{7-p}$ sphere on
the transverse $S^{8-p}$ (see figure 1). These configurations with
constant polar angle represent a D(8-p)-brane wrapped on a $S^{7-p}$
sphere and extended along the radial direction. These are the kind of
configurations we want to study in this paper. Actually, we will
consider first a more general situation in which the polar angle
depends only on the radial coordinate, \ie\ when
$\theta= \theta(r)$. It is a rather simple exercise to compute the
induced metric $g$ in this case. Moreover, by inspecting the form of
the RR potential $C^{(7-p)}$ in eq. (\ref{ocho}) and the Wess-Zumino
term in the action, one easily concludes that this term acts as a source
for the worldvolume electric field $F_{0,r}$ and, thus, it is natural
to assume that $F_{0,r}$ is different from zero. If we  take this
component of $F$ as the only non-vanishing one, the action can be
written as:
\beq
S\,=\,
\int_{S^{7-p}}\,d^{7-p}\theta\,\,
\int\,drdt\,\,{\cal L}(\theta, F)\,\,,
\label{trece}
\eeq
where the lagrangian density ${\cal L}(\theta, F)$ is given by:
\beq
{\cal L}(\theta, F)\,=\,-\,T_{8-p}\,R^{7-p}\,\sqrt{\hat g}\,\,\Bigl[\,
({\rm sin}\,\theta)^{7-p}
\,\,\sqrt{1\,+\,r^2\,\theta\,'^{\,2}\,-\,F_{0,r}^2}\,+\,
F_{0,r}\,C_p(\theta)\,\Bigr]\,\,.
\label{catorce}
\eeq
In eq. (\ref{catorce}) $\hat g$ is the determinant of the metric of the
$S^{7-p}$ and $\theta\,'$ denotes $d\theta/d r$. 

\medskip
\subsection{Quantization condition}
\medskip

The equation  of motion of the gauge field, derived from the
lagrangian density of eq. (\ref{catorce}), implies that:
\beq
{\partial {\cal L}\over \partial F_{0,r}}\,=\,
{\rm constant}\,\,.
\label{quince}
\eeq
In order to determine the value of the constant on the right-hand side
of eq. (\ref{quince}) let us follow the procedure of ref. \cite{PR}
and couple the D-brane to a Neveu-Schwarz (NS) Kalb-Ramond field $B$. As
is well-known, this coupling can be performed by substituting $F$ by
$F-B$ in ${\cal L}$, \ie\ by doing 
${\cal L}(\theta, F)\,\rightarrow {\cal L}(\theta, F-B)$ in eq.
(\ref{catorce}). At  first order in $B$, this substitution generates a
coupling of the D-brane to the NS field $B$ of the form:
\beq
\int_{S^{7-p}}\,\,d^{7-p}\theta\,\,\,\int\,drdt\,\,
{\partial \,{\cal L}\over\partial F_{0,r}}\,\,B_{0,r}\,\,,
\label{dseis}
\eeq
where we have assumed that only the $B_{0,r}$ component of the $B$
field is turned on. 

We shall regard eq. (\ref{dseis}) as the interaction energy of a
fundamental string source in the presence of the D-brane. This source 
is extended along the radial direction and, thus, it is quite natural
to require that the coefficient of the $B$ field, integrated over 
$S^{7-p}$, be an integer
multiple of the fundamental string tension, namely:
\beq
\int_{S^{7-p}}\,\,d^{7-p}\theta\,\,\,
{\partial \,{\cal L}\over\partial F_{0,r}}\,=\, n\,T_{f}\,\,,
\label{dsiete}
\eeq
with $n\in\ZZ$.  Eq. (\ref{dsiete}) is the quantization condition we
were looking for in these RR backgrounds and will play in our analysis
a role similar to the one played in ref. \cite{Bachas} by the flux
quantization condition (eq. (\ref{uno})). Notice that eq.  (\ref{dsiete})
constraints the electric components of $F$, whereas eq. (\ref{uno})
involves the magnetic worldvolume field\footnote{It is interesting to
point out that the left-hand side of eq. (\ref{dsiete}) can be written
in terms of the integral over the $S^{7-p}$ sphere of the worldvolume 
Hodge dual of $\partial \,{\cal L}/\partial 
F_{\alpha, \beta}$.}. Thus, our quantization rule
is a kind of electric-magnetic dual of the one used in ref.
\cite{Bachas}. This has a nice interpretation in the case in which $p$
is odd, which corresponds to the type IIB theory. Indeed, it is known in
this case  that the electric-magnetic
worldvolume duality corresponds to the S-duality of the
background \cite{EMduality}. In particular, when
$p=5$, the D5 background can be converted, by means of an S-duality
transformation,  into a NS5 one, which is precisely the type of
geometry considered in ref. \cite{Bachas}. 

By using the explicit form of the lagrangian density (eq.
(\ref{catorce})), the left-hand side of our quantization condition can
be easily calculated:
\beq
\int_{S^{7-p}}\,\,d^{7-p}\theta\,\,\,
{\partial \,{\cal L}\over\partial F_{0,r}}\,=\,
T_{8-p}\,\Omega_{7-p}\,R^{7-p}\,\Biggl[\,
{F_{0,r}\,\,({\rm sin}\,\theta)^{7-p}\,\over 
\sqrt{1\,+\,r^2\,\theta\,'^{\,2}\, -\,F_{0,r}^2}}\,\,-\,\,
C_p(\theta)\,\Biggr]\,\,,
\label{docho}
\eeq
where $\Omega_{7-p}$ is the volume of
the unit $(7-p)$-sphere, given by:
\beq
\Omega_{7-p}\,=\,{2\pi^{{8-p\over 2}}\over 
\Gamma\Bigl(\,{8-p\over 2}\Bigr)}\,\,.
\label{dnueve}
\eeq

By using eqs. (\ref{docho}) and (\ref{dsiete}) one can obtain 
$F_{0,r}$ as a function of $\theta(r)$ and the integer $n$. Let us
show how this can be done. First of all, by using eqs. 
(\ref{once}), (\ref{dnueve}) and (\ref{tres}) it is straightforward to
compute the global coefficient appearing on the right-hand side of eq. 
(\ref{docho}), namely:
\beq
T_{8-p}\,\Omega_{7-p}\,R^{7-p}\,=\,
{NT_f\over 2\sqrt{\pi}}\,
{\Gamma\Bigl(\,{7-p\over 2}\Bigr)\over
\Gamma\Bigl(\,{8-p\over 2}\Bigr)}\,\,. 
\label{veinte}
\eeq
Secondly, 
let us define the function ${\cal C}_{p,n}(\theta)$ as:
\beq
{\cal C}_{p,n}(\theta)\,=\,C_p(\theta)\,+\,2\,\sqrt{\pi}\,\,
{\Gamma\Bigl(\,{8-p\over 2}\Bigr)\over
\Gamma\Bigl(\,{7-p\over 2}\Bigr)}\,\,
{n\over N}\,\,.
\label{vuno}
\eeq
Notice that ${\cal C}_{p,n}(\theta)$ satisfies the same differential
equation as  $C_p(\theta)$ (eq. (\ref{seis})) with different initial
condition. Moreover,  by inspecting eqs. (\ref{dsiete}), (\ref{docho})
and  (\ref{veinte}), one easily concludes that $F_{0,r}$ can be put in
terms of ${\cal C}_{p,n}(\theta)$. The corresponding expression is:
\beq
F_{0,r}\,=\,\sqrt{\,
{1\,+\,r^2\,\theta\,'^{\,2}\over 
{\cal C}_{p,n}(\theta)^2\,+\,
({\rm sin}\,\theta)^{2(7-p)}}}\,\,{\cal C}_{p,n}(\theta)\,\,.
\label{vdos}
\eeq
Let us now evaluate the energy of the system. By performing a Legendre
transformation, we can represent the hamiltonian $H$ of the
D(8-p)-brane as:
\beq
H\,=\,\int_{S^{7-p}}\,\,d^{7-p}\theta\,\,\,
\int dr\,\Big[\, F_{0,r}\,{\partial \,{\cal L}\over\partial
F_{0,r}}\,-\, {\cal L}\,\Big]\,\,.
\label{vtres}
\eeq
By using (\ref{vdos}) one can eliminate $F_{0,r}$ from the expression
of $H$. One gets:
\beq
H\,=\,T_{8-p}\,\Omega_{7-p}\,R^{7-p}\,\int dr\,
\sqrt{1\,+\,r^2\,\theta\,'^{\,2}}\,\,
\sqrt{{\cal C}_{p,n}(\theta)^2\,+\,
({\rm sin}\,\theta)^{2(7-p)}}\,\,. 
\label{vcuatro}
\eeq
It is now simple to find the constant $\theta$ configurations which
minimize the energy. Indeed, we only have to require the vanishing of 
$\partial\,H/\partial\theta$ for $\theta^{'}=0$. Taking into account
that ${\cal C}_{p,n}(\theta)$ satisfies eq. (\ref{seis}), we arrive at:
\beq
{\partial\,H\over \partial\theta}\,\,\,\Biggr|_{\theta^{'}=0}\,\,\,=
\,\,\,(7-p)\,\,T_{8-p}\,\Omega_{7-p}\,R^{7-p}\,\,\,
{({\rm sin}\,\theta)^{7-p}\,
[\,({\rm sin}\,\theta)^{6-p}\,{\rm cos}\,\theta\,-\,
{\cal C}_{p,n}(\theta)\,\,]\over
\sqrt{{\cal C}_{p,n}(\theta)^2\,+\,
({\rm sin}\,\theta)^{2(7-p)}}}\,\,.
\label{vcinco}
\eeq
Moreover, if we define the function\footnote{The functions
$\Lambda_{p,n}(\theta)$ for different values of $p$ have been listed in
appendix A.}:
\beq
\Lambda_{p,n}(\theta)\,\equiv\,
({\rm sin}\,\theta)^{6-p}\,{\rm cos}\,\theta\,-\,
{\cal C}_{p,n}(\theta)\,\,,
\label{vseis}
\eeq
it is clear by looking at the right-hand side of eq. (\ref{vcinco}) 
 that the energy is minimized either when $\theta=0,\pi$ (\ie\ when 
$\sin\theta=0$) or when $\theta=\bar\theta_{p,n}$, where 
$\bar\theta_{p,n}$ is determined by the condition:
\beq
\Lambda_{p,n}(\bar\theta_{p,n})\,=\,0\,\,.
\label{vsiete}
\eeq
The solutions $\theta=0,\pi$ correspond to singular configurations in
which the D(8-p)-brane collapses at the poles of the $S^{7-p}$ sphere.
For this reason we shall concentrate on the analysis of the 
$\theta=\bar\theta_{p,n}$ configurations. First of all, we notice that
the function $\Lambda_{p,n}(\theta)$ has a simple derivative, which
can be obtained from its definition and from the differential equation
satisfied by ${\cal C}_{p,n}(\theta)$. One gets:
\beq
{d\over d\theta}\,\,\Lambda_{p,n}(\theta)\,=\,
(6-p)\,({\rm sin}\,\theta)^{5-p}\,\,.
\label{vocho}
\eeq
It follows from eq. (\ref{vocho}) that when 
$p< 6$ then ${d\over d\theta}\,\,\Lambda_{p,n}(\theta)\,>\, 0$ if
$\theta\in (0,\pi)$. This means that, for $p\le 5$,  
$\Lambda_{p,n}(\theta)$  is a monotonically increasing function in the
interval $0<\theta<\pi$. In what follows we shall
restrict ourselves to the case $p\le 5$. In order to check that eq. 
(\ref{vsiete}) has solutions in this case, let us evaluate the values
of $\Lambda_{p,n}(\theta)$ at $\theta=0,\pi$. From eqs.
(\ref{vseis}), (\ref{vuno})  and (\ref{siete}) we have: 
\beq
\Lambda_{p,n}(0)\,=\,-\,{\cal C}_{p,n}(0)\,=\,
-2\sqrt{\pi}\,\,
{\Gamma\Bigl(\,{8-p\over 2}\Bigr)\over
\Gamma\Bigl(\,{7-p\over 2}\Bigr)}\,\,
{n\over N}\,\,.
\label{vnueve}
\eeq
Moreover for $\theta=\pi$ we can write:
\beq
\Lambda_{p,n}(\pi)\,=\,-\,{\cal C}_{p,n}(\pi)\,=\,-\,
C_{p}(\pi)\,-\,2\sqrt{\pi}\,\,
{\Gamma\Bigl(\,{8-p\over 2}\Bigr)\over
\Gamma\Bigl(\,{7-p\over 2}\Bigr)}\,\,
{n\over N}\,\,,
\label{treinta}
\eeq
and,  taking into account that: 
\beq
C_{p}(\pi)\,=\,-(7-p)\,\int_{0}^{\pi}\,
({\rm sin}\,\theta)^{7-p}\,d\theta\,\,=\,\,-2\sqrt{\pi}\,\,
{\Gamma\Bigl(\,{8-p\over 2}\Bigr)\over
\Gamma\Bigl(\,{7-p\over 2}\Bigr)} \,\,,
\label{tuno}
\eeq
we get: 
\beq
\Lambda_{p,n}(\pi)\,=\,2\sqrt{\pi}\,\,
{\Gamma\Bigl(\,{8-p\over 2}\Bigr)\over
\Gamma\Bigl(\,{7-p\over 2}\Bigr)}\,\,
(\,1\,-\,{n\over N}\,)\,\,.
\label{tdos}
\eeq
As ${d\over d\theta}\,\,\Lambda_{p,n}(\theta)\,>\, 0$ for
$\theta\in (0,\pi)$, the function $\Lambda_{p,n}(\theta)$ vanishes for
$0<\theta <\pi$ if and only if $\Lambda_{p,n}(0)\,<0$ 
and $\Lambda_{p,n}(\pi)\,>\,0$. From eq. (\ref{vnueve}) we conclude
that the first condition occurs when $n>0$, whereas
eq. (\ref{tdos}) shows that 
$\Lambda_{p,n}(\pi)\,>\,0$ if $n<N$. It follows that there exists only
one solution $\bar\theta_{p,n}\in (0,\pi)$ of eq. (\ref{vsiete}) for
each
$n$ in the interval $0<n<N$. Then, we have found exactly $N-1$ 
angles which correspond to nonsingular wrappings of the D(8-p)-brane
on a $S^{7-p}$ sphere. Notice that for $n=0$ ($n=N$) the solution of
eq. (\ref{vsiete}) is $\bar\theta_{p,0}=0$ ($\bar\theta_{p,N}=\pi$) 
(see eqs. (\ref{vnueve}) and (\ref{tdos})). Therefore, we can identify
these $n=0,N$ cases with the singular configurations previously
found. In general, when $n$ is varied from $n=0$ to $n=N$ the angle 
 $\bar\theta_{p,n}$ increases from $0$ to $\pi$ (\ie\ from one of the
poles of the $S^{8-p}$ sphere to the other). 

It is not difficult to find the energy of these wrapped configurations.
Actually we only need to substitute $\theta=\bar\theta_{p,n}$ in 
eq. (\ref{vcuatro}). Taking into account (see eqs. (\ref{vseis}) and
(\ref{vsiete})) that:
\beq
{\cal C}_{p,n}(\bar\theta_{p,n})\,=\,
(\,\sin\bar\theta_{p,n}\,)^{6-p}\,\cos\bar\theta_{p,n}\,\,,
\label{ttres}
\eeq
one easily finds that the energy of these solutions can be written as:
\beq
H_{p,n}\,=\,\int\,dr\,{\cal E}_{p,n}\,\,,
\label{tcuatro}
\eeq
where the constant energy density ${\cal E}_{p,n}$ is given by:
\beq
{\cal E}_{p,n}\,=\,
{NT_f\over 2\sqrt{\pi}}\,
{\Gamma\Bigl(\,{7-p\over 2}\Bigr)\over
\Gamma\Bigl(\,{8-p\over 2}\Bigr)}\,\,
({\rm sin}\,\bar\theta_{p,n})^{6-p}\,\,.
\label{tcinco}
\eeq
Similarly, by substituting eq. (\ref{ttres}) in eq. (\ref{vdos}), we
can get the worldvolume electric field for our configurations, namely:
\beq
\bar F_{0,r}\,=\,{\rm cos}\,\bar\theta_{p,n}\,\,. 
\label{tseis}
\eeq
Let us now analyze some particular cases of our equations. First of
all, we shall consider the $p=5$ case, \ie\ a D3-brane wrapped on a
two-sphere under the action of a D5-brane background. The function
$\Lambda_{5,n}(\theta)$ is:
\beq
\Lambda_{5,n}(\theta)\,=\,\theta\,-\,{n\over N}\,\pi\,\,,
\label{tsiete}
\eeq
and the equation $\Lambda_{5,n}(\theta)=0$ is trivially solved by the
angles:
\beq
\bar\theta_{5,n}\,=\,{n\over N}\,\pi\,\,.
\label{tocho}
\eeq
Notice that the set of angles in eq. (\ref{tocho}) is the same as that
of ref. \cite{Bachas}. Using this result in eq. (\ref{tcinco}) we get the
following energy density:
\beq
{\cal E}_{5,n}\,=\,{NT_f\over \pi}\,
{\rm sin}\,\Big[\,{n\over N}\,\pi\,\Big]\,\,,
\label{tnueve}
\eeq
which is very similar to the result found in ref. \cite{Bachas}. Next,
let us take $p=4$, which corresponds to a D4-brane wrapped on a
three-sphere in the background of a stack of D4-branes. The corresponding 
$\Lambda_{p,n}(\theta)$ function is:
\beq
\Lambda_{4,n}(\theta)\,=\,-2\,\Big[\,{\rm cos}\,\theta\,+\,
2\,{n\over N}\,-\,1\,\Big]\,\,,
\label{cuarenta}
\eeq
and the solutions of eq. (\ref{vsiete}) in this case are easily found,
namely:
\beq
{\rm cos}\,\bar\theta_{4,n}\,=\,1\,-\,2\,{n\over N}\,\,.
\label{cuno}
\eeq
The corresponding energy density takes the form:
\beq
{\cal E}_{4,n}\,=\,{n(N-n)\over N}\,\,T_f\,\,.
\label{cdos}
\eeq
Notice that, in this D4-D4 case, the energy density of eq.
(\ref{cdos}) is a rational fraction of the fundamental string tension.

For general $p$ the equation $\Lambda_{p,n}(\theta)\,=\,0$ is much
difficult to solve analytically. In order to illustrate this point let
us write down the equation to solve in the physically interesting case
$p=3$:
\beq
\bar\theta_{3,n}\,-\,\cos\bar\theta_{3,n}\,\sin\,\bar\theta_{3,n}\,=\,
{n\over N}\,\pi\,\,.
\label{ctres}
\eeq
Despite of the fact that we are not able to find the analytical
solution of the equation $\Lambda_{p,n}(\theta)\,=\,0$ for $p\le 3$,
we can get some insight on the nature of our solutions from some general
considerations. First of all, it is interesting to point out the
following property of the functions $\Lambda_{p,n}(\theta)$:
\beq
\Lambda_{p,n}(\theta)\,=\,-\Lambda_{p,N-n}(\pi-\theta)\,\,.
\label{extrauno}
\eeq
Eq. (\ref{extrauno}) can be proved either from the definition of the 
$\Lambda_{p,n}(\theta)$'s or from their expressions listed in appendix A.
It follows from this equation that our set of angles 
$\bar\theta_{p,n}$ satisfy:
\beq
\bar\theta_{p,N-n}\,=\,\pi\,-\,\bar\theta_{p,n}\,\,.
\label{extrados}
\eeq
By using (\ref{extrados}) in the expression of the energy density 
${\cal E}_{p,n}$ (eq. (\ref{tcinco})), one immediately gets the
following periodicity relation:
\beq
{\cal E}_{p,N-n}\,=\,{\cal E}_{p,n}\,\,.
\label{extratres}
\eeq

Another interesting piece of information can be obtained by considering
the semiclassical $N\rightarrow\infty$ limit. Notice that 
$\Lambda_{p,n}$ depends on $n$ and $N$ through their ratio $n/N$ (see
eqs. (\ref{vseis})) and  (\ref{vuno})). Then, taking $N\rightarrow\infty$
with fixed $n$ is equivalent to make $n\rightarrow 0$ for finite $N$. We
have already argued that if $n\rightarrow 0$ the angle 
$\bar\theta_{p,n}\rightarrow 0$. In order to solve the equation 
$\Lambda_{p,n}(\theta)\,=\,0$ for small $\theta$, let us expand 
$\Lambda_{p,n}(\theta)$ in Taylor series around $\theta=0$. It turns
out \cite{Camino} that the first non-vanishing derivative of
$\Lambda_{p,n}(\theta)$ at $\theta=0$ is the $(6-p)^{th}$ one and,
actually, near $\theta=0$, we can write:
\beq
\Lambda_{p,n}(\theta)\,\approx\,\Lambda_{p,n}(0)\,+\,\theta^{6-p}\,
+\,\cdots\,\,.
\label{ccuatro}
\eeq
It follows immediately that for $N\rightarrow\infty$ the value of 
$\bar\theta_{p,n}$ is given by:
\beq
\big(\,\bar\theta_{p,n}\,\big)^{6-p}\,\approx\,-\Lambda_{p,n}(0)\,\,.
\label{ccinco}
\eeq
Taking into account eq. (\ref{vnueve}) and the general expression of
the energy density (eq. (\ref{tcinco})), we can easily verify that:
\beq
\lim_{N\rightarrow\infty}\,{\cal E}_{p,n}\,=\,n\,T_f\,\,,
\label{cseis}
\eeq
a fact which can be verified directly for $p=4,5$ from our analytical
expressions of the energy density (eqs. (\ref{tnueve}) and
(\ref{cdos})).  It is now clear from eq. (\ref{cseis}) that our
configurations can be interpreted as bound states of $n$ fundamental
strings. Actually, one can prove quite generally that the following
inequality holds:
\beq
{\cal E}_{p,n}\,\le\,n\,T_f\,\,,
\label{extracuatro}
\eeq
which shows that the formation of our bound states is energetically
favored. This is an indication of their stability, which we will
verify directly in section 2.3.

In order to prove (\ref{extracuatro}), it is  very useful again to
consider the dependence of the energy on $1/N$. Notice that for 
$1/N\rightarrow 0$ both sides of eq. (\ref{extracuatro}) are equal (see
eq. (\ref{cseis}) ). The energy ${\cal E}_{p,n}$ depends on $1/N$
both explicitly and implicitly  (through $\bar\theta_{p,n}$). If we
consider $1/N$ as a continuous variable, then one has:
\beq
{d\over d\Big({1\over N}\Big)}\,\,\bar\theta_{p,n}\,\,=\,\,
{nNT_f\over (6-p)\,{\cal E}_{p,n}}\,\,\sin\bar\theta_{p,n}\,\,.
\label{extracinco}
\eeq
Eq. (\ref{extracinco}) is obtained by differentiating eq. 
(\ref{vsiete}) and using eqs. (\ref{vocho}) and (\ref{vnueve}) (the
latter determines the explicit dependence of $\Lambda_{p,n}$ on 
$1/N$). We are now ready to demonstrate (\ref{extracuatro}). For this
purpose let us consider the quantity $({\cal E}_{p,n}-n\,T_f)/N$, which
we will regard as a function of $1/N$. We must prove that this quantity
is always less or equal than zero. Clearly,  eq. (\ref{cseis}) implies
that  $({\cal E}_{p,n}-n\,T_f)/N\rightarrow 0$ for
$1/N\rightarrow 0$. Moreover, by using (\ref{extracinco}) it is
straightforward to compute the derivative:
\beq
{d\over d\Big({1\over N}\Big)}\,\,\Bigg[\,
{{\cal E}_{p,n}-n\,T_f\over N}\,\Bigg]\,=\,-n\,T_f\,\,
(\,1\,-\,\cos\bar\theta_{p,n}\,)\,\,,
\label{extraseis}
\eeq
which vanishes for $N\rightarrow \infty$ and is always negative  for
finite $N$ and $0<n<N$. Thus, it follows that $({\cal E}_{p,n}-n\,T_f)/N$
is negative for finite $N$ and, necessarily, eq. (\ref{extracuatro})
holds. 

As a further check of (\ref{extracuatro}) one can compute the first
correction to ${\cal E}_{p,n}-n\,T_f$ for finite $N$. By Taylor
expanding ${\cal E}_{p,n}$ in powers of $1/N$, and using eq. 
(\ref{extracinco}), one can prove that:
\beq
{\cal E}_{p,n}\,\,-\,\,n\,T_f\,\approx\,\,-
\,{6-p\over 2(8-p)}\,\,n\,T_f\,\,\,\Big(\,
{\cal C}_{p,n}(0)\,\Big)^{{2\over 6-p}}\,\,+\,\cdots\,\,,
\label{extrasiete}
\eeq
where ${\cal C}_{p,n}(0)$, which is of order $1/N$, has been   given in
eq.  (\ref{vnueve}).

\medskip
\subsection{BPS configurations and the baryon vertex}
\medskip
In this section we shall show that the wrapped configurations found
above solve a BPS differential equation. With this purpose in mind,
let us now come back to the more general situation in which the angle
$\theta$ depends on the radial coordinate $r$. The hamiltonian for a
general function $\theta(r)$ was given in eq. (\ref{vcuatro}). By
means of a simple calculation it can be verified that this hamiltonian
can be written as:

\beq
H\,=\,T_{8-p}\,\Omega_{7-p}\,R^{7-p}\,\int dr\,
\sqrt{\,{\cal Z}^2\,+\,{\cal Y}^2}\,\,,
\label{csiete}
\eeq
where, for any function $\theta(r)$, ${\cal Z}$ is a total derivative:
\beq
{\cal Z}\,=\,{d\over dr}\,\Big[\,r\Big(\,
({\rm sin}\,\theta)^{6-p}\,-\,\Lambda_{p,n}(\theta\,
)\,{\rm cos}\,\theta\, \,\Big)\,\Big]\,\,,
\label{cocho}
\eeq
and ${\cal Y}$ is given by:
\beq
{\cal Y}\,=\,{\rm sin}\,\theta\,\Lambda_{p,n}(\theta\,)\,
-\,r\theta\,'\,\Big[\,({\rm sin}\,\theta)^{6-p}\,-\,
\Lambda_{p,n}(\theta\,)\,{\rm cos}\,\theta\,\Big]\,\,.
\label{cnueve}
\eeq
It follows from eq. (\ref{csiete}) that $H$ is bounded as:
\beq
H\,\ge\,T_{8-p}\,\Omega_{7-p}\,R^{7-p}\,\int dr\,
\big |\,{\cal Z}\,\big |\,\,.
\label{cincuenta}
\eeq
Since ${\cal Z}$ is a total derivative, the bound on the right-hand
side of eq. (\ref{cincuenta}) only depends on the boundary values of 
$\theta(r)$. This implies that any $\theta(r)$ saturating the bound is
also a solution of the equations of motion. This saturation of the
bound clearly occurs when ${\cal Y}\,=\,0$ or, taking into account eq. 
(\ref{cnueve}), when  $\theta(r)$ satisfies the following first-order
differential equation:
\beq
\theta\,'\,=\,{1\over r}\, \,\,
{{\rm sin}\,\theta\,\Lambda_{p,n}(\theta\,)\over
({\rm sin}\,\theta)^{6-p}\,-\,\Lambda_{p,n}(\theta\,)
\,{\rm cos}\,\theta}\,\,.
\label{ciuno}
\eeq
It is straightforward to verify directly that any solution $\theta(r)$
of eq. (\ref{ciuno}) also solves the second-order
differential equations of motion derived from the hamiltonian of eq. 
(\ref{vcuatro}). Moreover, by using eq. (\ref{ciuno}) to evaluate the
right-hand side of eq. (\ref{vdos}), one can demonstrate that the BPS
differential equation is equivalent to the following relation between
the electric field $F_{0,r}$ and $\theta(r)$:
\beq
F_{0,r}\,=\,\partial_r\,(\,r\,{\rm cos}\,\theta\,)\,=\,
{\rm cos}\,\theta\,-\,r\theta\,'\,{\rm sin}\,\theta\,\,.
\label{cidos}
\eeq
Notice now that eq. (\ref{ciuno}) admits solutions with 
$\theta={\rm constant}$ if and only if $\theta\,=\,0\,,\,\pi$ 
or when $\theta$ is a zero of 
$\Lambda_{p,n}(\theta\,)$. Thus our wrapped configurations are
certainly solutions of the BPS differential equation. As a
confirmation of this fact, let us point out that, for constant 
$\theta$, the electric field of eq. (\ref{cidos}) reduces to the value
displayed in eq. (\ref{tseis}). 

Eq. (\ref{ciuno}) was first proposed (for $p=3$) in ref. \cite{Imamura} to
describe the baryon vertex (see also refs. \cite{CGS, Craps, Camino})
\footnote{In these studies of the baryon vertex a different choice of
worldvolume coordinates is performed. Instead of taking  these
coordinates as in eq. (\ref{doce}), one takes 
$\sigma^{\alpha}\,=\,(\,t\,,\,
\theta^1\,,\,\cdots\,,\,\theta^{7-p}\,,\,\theta\,\,)$ and the embedding
of the D(8-p)-brane is described by a function
$r\,=\,r(\sigma^{\alpha})$.}. In ref. \cite{kappa} it was verified, by
looking at the $\kappa$-symmetry of the brane probe, that the condition 
(\ref{cidos}) is enough to preserve $1/4$ of the bulk supersymmetry.
Actually, following the results of ref. \cite{Camino}, it is not
difficult to obtain the general solution of the BPS differential equation 
(\ref{ciuno}). In implicit form this solution can be written as:
\beq
{[\,\Lambda_{p,n}(\theta\,)\,]^{{1\over 6-p}}\over
{\rm sin}\,\theta}\,=\,C\,r\,\,,
\label{citres}
\eeq
\begin{figure}
\centerline{\hskip -.8in \epsffile{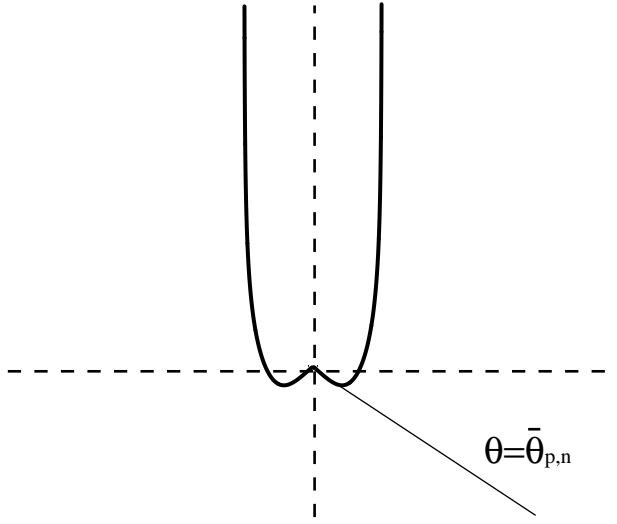}}
\caption{Representation of a typical solution of the BPS equation
(\ref{citres}) for $C\not= 0$. In this plot $r$ and $\theta$ are the
polar coordinates of the plane of the figure. We have also plotted the 
$\theta\,=\,\bar\theta_{p,n}$ curve, which  is the solution of
(\ref{citres}) for $C= 0$.}
\label{fig2}
\end{figure}
where $C$ is a constant. Our constant angle solutions
$\theta\,=\,\bar\theta_{p,n}$ can be obtained from eq.~(\ref{citres})
by taking $C=0$, whereas the baryon vertex solutions correspond to
$C\not= 0$. A glance at eq.~(\ref{citres}) reveals that, by
consistency, $\theta$ must be restricted to take values in an interval
such that the function $\Lambda_{p,n}(\theta\,)$ has a fixed sign. If,
for example, $\theta\in(0,\bar\theta_{p,n})$, then 
$\Lambda_{p,n}(\theta\,)<0$ and, by redefining the phase of $C$, we
get a consistent solution in which $r$ is a non-negative real number.
Similarly, we could have $\theta\in(\bar\theta_{p,n},\pi)$ since 
$\Lambda_{p,n}(\theta\,)>0$ for these values. In both cases 
$\bar\theta_{p,n}$ is a limiting angle. Actually, for 
$0<n<N$ one immediately infers from eq. (\ref{citres}) that 
$\bar\theta_{p,n}$ is the angle reached when $r\rightarrow 0$. The
baryon vertex solutions behave \cite{CGS,Craps, Camino} as a bundle of
fundamental strings in the asymptotic region $r\rightarrow \infty$ (see
figure 2). The number of fundamental strings is precisely $n$ for the
solution with 
$\theta\in(0,\bar\theta_{p,n})$ (and $N-n$ when 
$\theta\in(\bar\theta_{p,n},\pi)\,$). Notice that $r\rightarrow
\infty$ when $\theta=0$ ($\theta=\pi$) for the solution with $n$
($N-n$) fundamental strings, whereas in the opposite limit 
$r\rightarrow 0$ the solution displayed in eq. (\ref{citres}) is
equivalent to our $\theta\,=\,\bar\theta_{p,n}$  wrapped
configuration. This is quite suggestive and implies that one can
regard our constant angle configurations as a short distance limit (in
the radial direction) of the baryon vertex solutions.

\medskip
\subsection{Fluctuations and stability}
\medskip
We are now going to study fluctuations around the static
configurations found above. Let us parametrize these fluctuations as
follows: 
\beq
\theta\,=\,\bar\theta_{p,n}\,+\,\xi\,\,,
\,\,\,\,\,\,\,\,\,\,\,\,\,\,\,\,\,\,\,\,
F_{0,r}\,=\,\cos\bar\theta_{p,n}\,+\,f\,\,,
\label{cicuatro}
\eeq
where $\xi$ and $f$ are small quantities which depend on the worldvolume
coordinates $\sigma^{\alpha}$. We are going to prove in this section
that the $\theta=\bar\theta_{p,n}$ solution is stable under the
perturbation of eq. (\ref{cicuatro}). In order to achieve this goal we
must go back to the action written in eq. (\ref{diez}).  We shall
evaluate this action for an angle $\theta$ and an electric field as in
eq. (\ref{cicuatro}). Let us represent the perturbation $f$
by means of a potential as $f\,=\,\partial_0a_r\,-\,\partial_ra_0$.
We shall choose a gauge in which the components $a_{\hat i\,}$ of the
potential  along the sphere $S^{7-p}$ vanish. 
Then we see that, for consistency, we  must include in our
perturbation the components of the gauge field  strength of the type 
$F_{\hat  i\,,r}\,=\,\partial_{\hat i\,}a_r$ and
$F_{0,\hat  i}\,=\,-\partial_{\hat i\,}a_0$. Under these circumstances
it is not difficult to compute the lagrangian density for the action
(\ref{diez}) up to second order in $\xi$, $f$,  $F_{\hat  i\,,r}$
and $F_{0,\hat  i}$.
After some calculation one gets:
\bear
{\cal L}\,&=&\,-\sqrt {\hat g}\,R^{7-p}\,T_{8-p}\,
\Lambda_{p,n}(0)\,f\,+\,\sqrt {\hat g}\,R^{7-p}\,T_{8-p}\,
(\,\sin\bar\theta\,)^{6-p}\,\times\rc\rc
&\times&{1\over 2}\Bigg\{\,
R^{7-p}\,r^{p-5}\,(\,\partial_0\xi\,)^2\,-\, 
r^2\,(\,\partial_r\xi\,)^2\,-\,(\,\partial_{\hat i\,}
\xi\,)^2\,+\,\rc\rc  
&+&{R^{p-7}\,r^{5-p}\over
(\sin\bar\theta)^2}\,\big[\,\Bigg({R\over r}\Bigg)^{7-p}\,F_{0,\,\hat
i}^2\,-\, F_{\hat i,\,r}^2\,\big]\,+\,
\,(7-p)\xi^2\,+\,{f^2\over (\sin\bar\theta)^2}\,+\,
2(7-p)\,\,{f\xi\over \sin\bar\theta}\,\,\Bigg\}\,\,,\rc
\label{cicinco}
\eear
where, to simplify the notation, we have written 
$\bar\theta$ instead of $\bar\theta_{p,n}\,$, $\hat g_{\hat i\,\hat j}$
represents the metric of the $S^{7-p}$ sphere and we have denoted:
\beq
(\,\partial_{\hat i\,}\xi\,)^2\,=\,\hat
g^{\hat i\,\hat j}\partial_{\hat i\,}\xi\,
\partial_{\hat j\,}\xi\,\,,
\,\,\,\,\,\,\,\,\,\,\,\,\,\,\,
F_{\hat i,\,r}^2\,=\,\hat g^{\hat i\hat j}F_{\hat i,\,r}\,F_{\hat j,\,r}
\,\,,
\,\,\,\,\,\,\,\,\,\,\,\,\,\,\,
F_{0\,,\hat i}^2\,=\,\hat g^{\hat i\hat j}F_{0\,,\hat i}
\,F_{0\,,\hat j}
\,\,.
\label{ciseis}
\eeq
In eq. (\ref{cicinco}) we have dropped the zero-order term. 
Moreover, the first term on the right-hand side of eq.
(\ref{cicinco}) is a first-order term which, however, does not
contribute to the equations of motion. In fact, by computing the
variation of the action with respect to $a_0$, $a_r$ and $\xi$ we get
the following set of equations:
\bear
&&\partial_r\,\Big[\,{f\over \sin\bar\theta}\,
+\,(7-p)\,\xi\,\Big]\,+\,
{1\over r^2\,\sqrt{\hat g}}\,
\partial_{\hat i\,}\,\Big[\,\sqrt{\hat g}\,\hat g^{\hat i\,\hat j}\,
{F_{0,\hat j}\over \sin\bar\theta}\,\Big]\,
\,=\,0\,\,,\rc\rc
&&r^{p-5}\,R^{7-p}\partial_{0}\,\Big[\,{f\over \sin\bar\theta}\,
+\,(7-p)\,\xi\,\Big]\,-\,{1\over \sqrt{\hat g}}\,
\partial_{\hat i\,}\,\Big[\,\sqrt{\hat g}\,\hat g^{\hat i\,\hat j}\,
{F_{\hat j,\,r}\over \sin\bar\theta}\,\Big]\,=\,0\,\,,\rc\rc
&&R^{7-p}r^{p-5}\,\partial_0^2\xi\,-\,
\partial_r(r^2\partial_r\xi)\,-\,\nabla^2_{S^{(7-p)}}\,\xi\,+\,
(p-7)\,\Big[\,\xi\,+\,{f\over \sin\bar\theta}\,\Big]\,=\,0\,\,.\rc
\label{cisiete}
\eear
The first equation in (\ref{cisiete}) is nothing but the Gauss law.
Moreover,  if we further fix the gauge to $a_0=0$ 
(\ie\ $f\,=\,\partial_0a_r$, 
$F_{\hat i,\,r}\,=\,\partial_{\hat i\,}a_r$ and 
$F_{0,\hat  i}\,=\,0$), the second equation  in (\ref{cisiete}) can be
written as:
\beq
r^{p-5}\,R^{7-p}\,\Big[\,{\partial_{0}^2a_r\over \sin\bar\theta}\,+\,
(7-p)\,\partial_{0}\xi\,\Big]\,-\,{1\over \sin\bar\theta}\,
\nabla^2_{S^{(7-p)}}\,a_r\,=\,0\,\,,
\label{ciocho}
\eeq
where $\nabla^2_{S^{(7-p)}}$ is the laplacian operator on the 
$S^{(7-p)}$ sphere. In order to continue with our analysis,
let us now expand $a_r$ and $\xi$ in spherical harmonics of
$S^{(7-p)}$:
\bear
&&a_r(\,t,r,\theta^1,\cdots,\theta^{7-p}\,)\,=\,
\sum_{l\ge 0, m}\,Y_{l,m}(\,\theta^1,\cdots,\theta^{7-p}\,)\,
\alpha_{l,m}(\,t,r\,)\,\,,\rc\rc
&&
\xi(\,t,r,\theta^1,\cdots,\theta^{7-p}\,)\,=\,
\sum_{l\ge 0, m}\,Y_{l,m}(\,\theta^1,\cdots,\theta^{7-p}\,)\,
\zeta_{l,m}(\,t,r\,)\,\,.\rc
\label{cinueve}
\eear
The spherical harmonics $Y_{l,m}$ are well-defined functions on 
$S^{(7-p)}$ which are eigenfunctions of the laplacian on the sphere,
namely:
\beq
\nabla^2_{S^{(7-p)}}\,Y_{l,m}\,=\,
-l(l+6-p)\,Y_{l,m}\,\,.
\label{sesenta}
\eeq
By plugging the mode expansion (\ref{cinueve}) into the equations of
motion (\ref{cisiete}) and (\ref{ciocho}), and using eq.
(\ref{sesenta}), we can obtain some equations for 
$\alpha_{l,m}(\,t,r\,)$ and $\zeta_{l,m}(\,t,r\,)$. Actually, if we
define:
\beq
\eta_{l,m}\,\equiv\,
{\partial_{0}\alpha_{l,m}\over\sin\bar\theta}\,+\,(7-p)\,\zeta_{l,m}
\,\,,
\label{suno}
\eeq
then, the Gauss law in this $a_0\,=\,a_{\hat i}\,=\,0$ gauge can be
simply written as:
\beq
\partial_{r}\,\eta_{l,m}\,=\,0\,\,,
\label{sdos}
\eeq
whereas the other two equations of motion give rise to:
\bear
&&R^{7-p}\,r^{p-5}\,\partial_0\,\Big[\,
{\partial_{0}\alpha_{l,m}\over\sin\bar\theta}\,+\,(7-p)\,\zeta_{l,m}
\,\Big]\,+\,l(l+6-p)\, {\alpha_{l,m}\over\sin\bar\theta}\,=\,0\,\,,
\rc\rc
&&R^{7-p}\,r^{p-5}\,\partial_0^2\zeta_{l,m}\,-\,
\partial_r(\,r^2\,\partial_r\,\zeta_{l,m})\,
+\,l(l+6-p)\,\zeta_{l,m}\,+\rc\rc
&&+\,(p-7)\,\Big[\,\zeta_{l,m}\,+
{\partial_{0}\alpha_{l,m}\over\sin\bar\theta}\,\Big]\,=\,0\,\,.
\rc
\label{stres}
\eear
Let us analyze first eqs. (\ref{sdos}) and (\ref{stres}) for 
$l=0$. From the first equation in (\ref{stres}) it follows that:
\beq
\partial_0\eta_{0,m}\,=\,0\,\,.
\label{scuatro}
\eeq
Thus,  as $\partial_r\eta_{0,m}\,=\,0$ (see eq. (\ref{sdos})), one
concludes that:
\beq
\eta_{0,m}\,=\,{\rm constant}\,\,.
\label{scinco}
\eeq
By using this result and the definition of $\eta_{l,m}$ given in eq. 
(\ref{suno}), we can express $\partial_0\,\alpha_{0,m}$ in terms of 
$\zeta_{0,m}$ and the additive constant appearing in eq. 
(\ref{scinco}). By substituting this relation in the second equation
in (\ref{stres}), we get:
\beq
R^{7-p}\,r^{p-5}\,\partial_0^2\zeta_{0,m}\,-\,
\partial_r(\,r^2\,\partial_r\,\zeta_{0,m})\,+\,
(6-p)(7-p)\,\zeta_{0,m}\,=\,{\rm constant}\,\,.
\label{sseis}
\eeq
It is interesting to rewrite eq. (\ref{sseis}) in the following form.
First of all, we define the wave operator ${\cal O}_p$ that acts on any
function $\psi$ as:
\beq
{\cal O}_p\,\psi\equiv\,R^{7-p}\,r^{p-5}\,\partial_0^2\,\psi\,-\,
\partial_r\,(\,r^2\,\partial_r\psi\,)\,\,.
\label{ssiete}
\eeq
Then, if $m^2_0$ is given by:
\beq
m^2_0\,=\,(6-p)(7-p)\,\,,
\label{socho}
\eeq
eq. (\ref{sseis}) can be written as:
\beq
\Big(\,\,{\cal O}_p\,+\,m^2_0\,\,\Big)\,\zeta_{0,m}\,
=\,{\rm constant}\,\,,
\label{snueve}
\eeq
which means that $\zeta_{0,m}$ is a massive mode with mass $m_0$.
Notice that, as $p<6$, $m^2_0$ is strictly positive.

For a general value of $l>0$ the equations of motion can be
conveniently expressed in terms of the variables  $\eta_{l,m}$ and
$\zeta_{l,m}$. Indeed, by differentiating with respect to the time the
first equation (\ref{stres}), and using the definition (\ref{suno}),
we can put them in terms of  $\eta_{l,m}$ and
$\zeta_{l,m}$. Actually, if we define the mass matrix ${\cal M}_p$ as:
\beq
{\cal M}_p\,=\,\pmatrix{
l\,(l+6-p)\,+\,(7-p)\,(6-p)&&p-7\cr\cr
(p-7)\,l\,(l+6-p)&&l\,(l+6-p)}\,\,,
\label{setenta}
\eeq
the equations of motion can be written as:
\beq
\Big(\,\,{\cal O}_p\,+\,{\cal M}_p\,\,\Big)\,
\pmatrix{\zeta_{l,m}\cr\eta_{l,m}}
=\,0\,\,,
\label{stuno}
\eeq
where ${\cal O}_p$ is the wave operator defined in eq. (\ref{ssiete}).
In order to check that our wrapped configurations are stable, we must
verify that the eigenvalues of the matrix ${\cal M}_p$ are
non-negative. After a simple calculation one can show that these
eigenvalues are:
\beq
m_l^2\,=\,\cases{(l+6-p)\,(l+7-p)&for $l=0,1,\cdots \,\,,$\cr\cr
              l\,(l-1)&for $l=1,2,\cdots\,\,,$\cr}
\label{stdos}
\eeq
where we have already included the $l=0$ case. Eq. (\ref{stdos})
proves that there are not negative mass  modes in the spectrum of
small oscillations for $p<6$, which demonstrates that, as claimed, our
static solutions are stable.

\setcounter{equation}{0}
\section{Flux stabilization of  M5-branes}
\medskip

In this section we are going to describe a mechanism of flux
stabilization in M-theory. We shall consider a particular solution of
the equations of motion of eleven dimensional supergravity which is
the one associated to a stack of $N$ parallel M5-branes. The metric of
this solution takes the form \cite{supergravity}:
\beq
ds^2\,=\,{r\over R}\,(-dt^2\,+\,dx^1\,+\cdots +\,dx_5^2\,)\,+\,
{R^2\over r^2}\,(dr^2\,+\,r^2\,d\Omega_4^2\,)\,\,,
\label{sttres}
\eeq
where the ``radius" $R$ is given by:
\beq
R^3\,=\,\pi\, N\,l_p^3\,\,.
\label{stcuatro}
\eeq
In eq. (\ref{stcuatro}) $l_p$ is the Planck length in eleven
dimensions. The M5-brane solution of D=11 supergravity has also a
non-vanishing value of the four-form field strength $F^{(4)}$, under
which the M5-branes are magnetically charged. This field strength
is given by:
\beq
F^{(4)}\,=\,3\,R^3\,\epsilon_{(4)}\,\,.
\label{stcinco}
\eeq 
It is not difficult to find a three-form potential $C^{(3)}$ such that
$F^{(4)}\,=\,dC^{(3)}$. Actually, if we decompose the $S^4$ line
element $d\Omega_4^2$ as in eq. (\ref{cinco}) and use the same
orientation conventions as in section 2,  one can readily check that 
$C^{(3)}$ can be taken as:
\beq
C^{(3)}\,=\,-R^3\,C_4(\theta)\epsilon_{(3 )}\,\,,
\label{stseis}
\eeq
where $C_4(\theta)$ is the function
defined in  eqs. (\ref{seis}) and (\ref{siete}), namely 
$C_4(\theta)\,=\,\cos\theta\sin^2\theta\,+\,2(\cos\theta\,-\,1)$.

We will put in this background a probe M5-brane, whose action will be
given by the so-called PST formalism \cite{PST}. The fields of this
formalism include a three-form field strength $F$, whose potential is a
two-form field $A$ (\ie\ $F=dA$) and a scalar field $a$ (the PST scalar).
The field strength $F$ can be combined with (the pullback of) the
background potential $C^{(3)}$ to form the field
\footnote{We hope that this field $H$ will not be confused with the
hamiltonian.} $H$ as:
\beq
H\,=\,F\,-\,C^{(3)}\,\,.
\label{stsiete}
\eeq
Let us now define the field ${\tilde H}$ as follows:
\beq
{\tilde H}^{ij}\,=\,{1\over 3!\,\sqrt{-{\rm det}\,g}}\,
{1\over \sqrt{-(\partial a)^2}}\,
\epsilon^{ijklmn}\,\partial_k\,a\,H_{lmn}\,\,,
\label{stocho}
\eeq
where $g$ is the induced metric on the M5-brane worldvolume. 
The PST action of the M5-brane probe is:
\bear
S\,&=&\,T_{M5}\,\int\,d^6\sigma\,
\Bigg[\,-\sqrt{-{\rm det}(g\,+\,\tilde H)}\,+\,
{\sqrt{-{\rm det}g}\over 4\partial a\cdot\partial a}\,
\partial_i a\,(^*H)^{ijk}\,H_{jkl}\partial^l a\,
\Bigg]\,+\rc\rc
&&+\,T_{M5}\int\Bigg[\,C^{(6)}\,
+\,{1\over 2}\,F\wedge\,C^{(3)}\,\Bigg]\,\,,
\label{stnueve}
\eear
where $^*H$ denotes the Hodge dual of $H$, $C^{(6)}$ is (the pullback
of) the 6-form potential dual to $C^{(3)}$, and the M5-brane tension is
given by:
\beq
T_{M5}\,=\,{1\over (2\pi)^5\,l_p^6}\,\,.
\label{ochenta}
\eeq

We will extend our M5-brane probe along the  directions transverse to
the M5-branes of the background and along one of the directions
parallel to them. Without loss of generality we will take the latter to
be the $x^5$ direction. Accordingly, our worldvolume coordinates
$\sigma^{\alpha}$ will be taken to be:
\beq
\sigma^{\alpha}\,=\,(\,t,r,x^5,\theta^1,\theta^2,\theta^3)\,\,,
\label{ouno}
\eeq
and the embedding of the M5-brane probe is determined by a function
$\theta\,=\,\theta(\,\sigma^{\alpha}\,)$. As in the case of the RR
background, we shall mainly look for solutions with 
$\theta\,=\,{\rm constant}$, which represent a M5-brane wrapped on a
three-sphere and extended along the $r$ and $x^5$ directions. 

As discussed in ref. \cite{PST}, the scalar  $a$ is an auxiliary
field which can be eliminated from the action by fixing its  gauge
symmetry. The price one must pay for this gauge fixing is the loss of
manifest covariance. A particularly convenient choice for $a$  is:
\beq
a\,=\,x^5\,\,.
\label{odos}
\eeq
In this gauge the components of the worldvolume potential $A$ with
$x^5$ as one of its indices can be gauge fixed to zero \cite{PST}.
Moreover, if we consider configurations of $A$ and of the embedding angle
$\theta$ which are independent of $x^5$, one readily realizes that the
components of the three-forms $F$ and $H$ along $x^5$ also vanish and,
as  a consequence, only the square root term of the PST action 
(\ref{stnueve}) is non-vanishing. As we will verify soon this
constitutes a great simplification.

\medskip
\subsection{Quantization condition and M5-brane configurations}
\medskip

In order to find stable $S^3$-wrapped configurations of the M5-brane
probe, we need to switch on a non-vanishing worldvolume field which
could prevent the collapse to one of the poles of the $S^3$. As in
section 2 (and ref. \cite{Bachas}) the value of this worldvolume field
is determined by some quantization condition which can be obtained by
coupling the M5-brane to an open M2-brane. 

\begin{figure}
\centerline{\hskip -.8in \epsffile{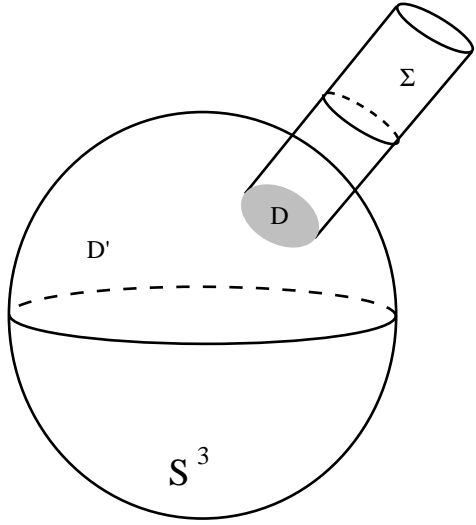}}
\caption{An M2-brane with worldvolume $\Sigma$ having its boundary on
the worldvolume of an M5-brane. If $\Sigma$ is attached to a submanifold
of the M5-brane worldvolume with the topology of $S^3$, there are two
possible disks $D$ and $D'$ on the $S^3$ whose boundary is $\partial
\Sigma$.}
\label{fig3}
\end{figure}

Let us consider an open M2-brane with worldvolume given by a
three-manifold $\Sigma$ whose boundary $\partial\Sigma$ 
lies on the worldvolume of an
M5-brane. For simplicity we shall consider the  case 
in which  $\partial\Sigma$ has only one
component (see figure 3). Clearly, $\partial\Sigma$ is also the boundary
of some disk $D$ on the worldvolume of the M5-brane. Let $\hat\Sigma$ be a
four-manifold whose boundaries are $\Sigma$ and $D$, \ie\ 
$\partial\hat\Sigma\,=\,\Sigma\,+\,D$. The coupling of the brane
to the supergravity background and to the M5-brane is described by an
action of the form:

\beq
S_{int}\,[\,\hat\Sigma, D\,]\,=\,
T_{M2}\,\int_{\hat\Sigma}\,F^{(4)}\,+\,T_{M2}\,\int_{D}\,H\,\,,
\label{otres}
\eeq
where $T_{M2}$ is the tension of the M2-brane, given by:
\beq
T_{M2}\,=\,{1\over (2\pi)^2\,l_p^3}\,\,.
\label{ocuatro}
\eeq
In a topologically trivial situation, if we represent $F^{(4)}$ as 
$dC^{(3)}$ and $F=dA$,  the above action reduces to the more familiar
expression:
\beq
S_{int}\,=\,T_{M2}\,\int_{\Sigma}\,C^{(3)}\,+
\,T_{M2}\,\int_{\partial D}\,A\,\,.
\label{ocinco}
\eeq

We shall regard eq. (\ref{otres}) as the definition of the interaction
term of the M2-brane action. Notice that, in general, $\hat\Sigma$ and
$D$ are not uniquely defined. To illustrate this point let us consider
the case in which we attach the M2-brane to a M5-brane worldvolume which
has some  submanifold with the topology of $S^3$. This is precisely the
situation in which we are interested in. As illustrated in figure 3, we
have two possible elections for the disk in eq. (\ref{otres}) namely,
we can choose the ``internal" disk $D$ or the ``external" disk
$D'$. Changing $D\rightarrow D'$, the manifold $\hat\Sigma$ changes to 
$\hat\Sigma'$, with $\partial \hat\Sigma'\,=\,\Sigma\,+\,D'$ and, in
general, $S_{int}$ also changes. However, in the quantum-mechanical
theory, the action appears in a complex exponential of the form 
$\exp[\,i\,S_{int}\,]$. Thus, we should require that:
\beq
e^{iS_{int}\,[\,\hat\Sigma, D\,]}\,=\,
e^{iS_{int}\,[\,\hat\Sigma', D'\,]}\,\,.
\label{oseis}
\eeq
The condition (\ref{oseis}) is clearly equivalent to:
\beq
S_{int}\,[\,\hat\Sigma', D'\,]\,-\,S_{int}\,[\,\hat\Sigma, D\,]\,=\,
2\pi n\,\,,
\label{osiete}
\eeq
with $n\in\ZZ$. The left-hand side of eq. (\ref{osiete}) can be
straightforwardly computed from eq. (\ref{otres}). Actually, 
if $\hat{\cal B}$ is the 4-ball bounded by $D'\cup(-D)\,=\,S^3$,
one has:
\beq
S_{int}\,[\,\hat\Sigma', D'\,]\,-\,S_{int}\,[\,\hat\Sigma, D\,]\,=\,
T_{M2}\,\int_{\hat{\cal B}}\,F^{(4)}\,+\,T_{M2}\,
\int_{\partial\hat{\cal B}}\,H\,\,.
\label{oocho}
\eeq
Using this result in eq. (\ref{osiete}), we get the condition:
\beq
\int_{\hat{\cal B}}\,F^{(4)}\,+\,
\int_{\partial\hat{\cal B}}\,H\,=\,{2\pi n\over T_{M2}}\,,
\,\,\,\,\,\,\,\,\,\,\,\,\,\,\,\,\,
n\in\ZZ\,\,.
\label{onueve}
\eeq
If $F^{(4)}$ can be represented as $dC^{(3)}$ on $\hat{\cal B}$, the
first integral on the left-hand side of eq. (\ref{onueve})
can be written as an integral of $C^{(3)}$ over 
$\partial\hat{\cal B}\,=\,S^3$. Our parametrization of $C^{(3)}$ 
(eq. (\ref{stseis})) is certainly non-singular if we  are
outside of the poles of the $S^4$. If this is  the case we get the
quantization condition:
\beq
\int_{S^3}\,F\,=\,{2\pi n\over T_{M2}}\,\,,
\,\,\,\,\,\,\,\,\,\,\,\,\,\,\,\,\,
n\in\ZZ\,\,.
\label{noventa}
\eeq
Eq. (\ref{noventa}), which is the M-theory analogue of eq. 
(\ref{uno}), is the quantization condition we were looking for. It is 
very simple  to obtain a solution of this equation. Let us take
$F$ proportional to the volume element $\epsilon_{(3)}$ of  the $S^3$.
Taking into account that the volume of the unit three-sphere is
$\Omega_3\,=\,2\pi^2$ (see eq. (\ref{dnueve})), we can write down
immediately the following solution of eq. (\ref{noventa}):
\beq
F\,=\,{n\over \pi T_{M2}}\,\,\epsilon_{(3)}\,\,.
\label{nuno}
\eeq
We can put this solution in a more convenient form if we use the
following relation between the M2-brane tension and the radius $R$:
\beq
T_{M2}\,=\,{N\over 4\pi R^3}\,\,, 
\label{ndos}
\eeq
which follows from eqs. (\ref{stcuatro}) and (\ref{ocuatro}). By using 
eq. (\ref{ndos}), one can rewrite eq. (\ref{nuno}) as:
\beq
F\,=\,4R^3\,{n\over N}\,\epsilon_{(3)}\,\,.
\label{ntres}
\eeq
We can  use the ansatz (\ref{ntres}) and the potential $C^{(3)}$ of
eq. (\ref{stseis}) to compute the three-form field $H$ of eq. 
(\ref{stsiete}). It turns out that the result for $H$ can be written
in terms of the function ${\cal C}_{4,n}(\theta)$ defined in eq. 
(\ref{vuno}). One gets:
\beq
H\,=\,R^3\,{\cal C}_{4,n}(\theta)\,\epsilon_{(3)}\,\,.
\label{ncuatro}
\eeq

Let us now assume that the angle $\theta$ characterizing the M5-brane
embedding only depends on the radial coordinate $r$ and, as
before, let us denote by $\theta'$ to the derivative $d\theta/dr$. As
was mentioned above, in the $a=x^5$ gauge and for this kind of
embedding, only the first term of the PST action (\ref{stnueve}) is
non vanishing and, as a consequence, all the dependence on $H$ of this
action comes through the field $\tilde H$ defined in eq.
(\ref{stocho}). Actually, the only non-vanishing component of
$\tilde H$ is:
\beq
\tilde H_{0r}\,=\,-{i\over (\sin\theta)^3}\,
\sqrt{{R\over r}}\,\sqrt{1\,+\,r^2\theta'^2}\,\,
{\cal C}_{4,n}\,(\theta)\,\,.
\label{ncinco}
\eeq
After a simple calculation one can obtain the induced metric $g$ and,
using eq. (\ref{ncinco}), the lagrangian density of the M5-brane. The
result is:
\beq
{\cal L}\,=\,-T_{M5}\,R^3\,\sqrt{\hat g}\,\,
\sqrt{1\,+\,r^2\theta'^2}\,
\sqrt{\,(\sin\theta)^6\,+\,(\,{\cal C}_{4,n}(\theta)\,)^2}\,\,,
\label{nseis}
\eeq
where $\hat g$ is the determinant of the metric of a unit 3-sphere. 
Notice the close similarity of this result and the hamiltonian density
of eq. (\ref{vcuatro}) for $p=4$, \ie\ for the D4-D4 system. Indeed,
it is immediate to check that the solutions with constant $\theta$ are
the same in both systems, \ie\ $\theta=\bar\theta_{4,n}$ with $0<n<N$,
where  $\bar\theta_{4,n}$ is given in eq. (\ref{cuno}) (for $n=0,N$
we have the singular solutions with $\theta=0,\pi$). This result is
quite natural since the D4-D4 system can be obtained from the M5-M5
one by means of a double dimensional reduction along the $x^5$
direction. The energy density for these solutions can be easily
obtained from the lagrangian (\ref{nseis}). One gets:
\beq
{\cal E}_n^{M5}\,=\,{n(N-n)\over N}\,T_{M2}\,\,,
\label{nsiete}
\eeq
which, again, closely resembles   the D4-D4 energy of eq.
(\ref{cdos}). In particular ${\cal E}_n^{M5}\rightarrow n\,T_{M2}$ as
$N\rightarrow\infty$, which implies that, semiclassically, our
configurations can be regarded as bound states of M2-branes. Moreover,
one can check that eq. (\ref{ciuno}) with $p=4$ is a BPS condition for
the M5-brane system. The integration of this equation can be read from
eq.  (\ref{citres}) and represents a baryonic vertex in M-theory
\cite{kappa}, $n$ being the number of M2-branes which form the baryon at
$r\rightarrow\infty$. The $\theta=\bar\theta_{4,n}$ solution can be
obtained as the $r\rightarrow 0$ limit of the M-theory baryon, in
complete analogy with the analysis at the end of section 2.2.

\medskip
\subsection{Fluctuations and stability}
\medskip
We will now perturb our static solution in order to check its
stability. As in section 2.3, we must allow the angle $\theta$ to
deviate from $\bar\theta_{4,n}$ and the worldvolume field strength
$F$ to vary from the value displayed in eq. (\ref{ntres}). The best
way to find out which components of $F$ must be included in the 
perturbation is to choose a gauge. As  $F$ in eq. (\ref{ntres}) has
only components along the sphere $S^3$, one can represent it by means
of a potential $\bar A_{\hat i\,\hat j}$ which also has component only
on $S^3$ (in what follows indices along $S^3$  will be denoted with a
hat). Accordingly, the perturbation of $F$ will be parametrized as a
fluctuation of the $S^3$-components of the potential $A$. Thus, we put:
\beq
\theta\,=\,\bar\theta_{4,n}\,+\,\xi\,\,,
\,\,\,\,\,\,\,\,\,\,\,\,\,\,\,\,\,\,\,\,
A_{\hat i\,\hat j}\,=\,\bar A_{\hat i\,\hat j}\,
+\,\alpha_{\hat i\,\hat j}\,\,,
\label{nocho}
\eeq
where $\xi$ and $\alpha_{\hat i\,\hat j}$ are small. For simplicity 
we shall assume that $\xi$ and the $\alpha_{\hat i\,\hat j}$'s do
not depend on $x^5$. Using the parametrization of $A$ in eq. 
(\ref{nocho}), it is clear that the $S^3$-components of the three-form
field $H$ can be written as:
\beq
H_{\hat i\,\hat j\,\hat k}\,=\,R^3\,
[\,{\cal C}_{4,n}(\theta)\,+\,f\,]\,\,
\,\,\,{\epsilon_{\hat i\,\hat j\,\hat k}\over \sqrt{\hat g}}\,\,,
\label{nnueve}
\eeq
where $f$ can be put in terms of derivatives of the type 
$\partial_{\hat i}\,\alpha_{\hat j\,\hat k}$. In eq. (\ref{nnueve})
$\hat g$ is the determinant of the metric of the $S^3$ and we are
using the convention 
$\epsilon^{\hat1\, \hat2\,\hat 3\,}\,=\,
\epsilon_{\hat 1\, \hat 2\,\hat 3\,}/{\hat g}\,=\,1$. As 
$\alpha_{\hat i\,\hat j}$ in (\ref{nocho}) depends on $t$ and $r$, it
follows that we have now
non-zero components 
$H_{0\hat i\,\hat j}\,=\,\partial_0\,\alpha_{\hat i\,\hat j}$ 
and $H_{r\hat i\,\hat j}\,=\,\partial_r\,\alpha_{\hat i\,\hat j}$.
Thus, in the gauge (\ref{odos}), the non-vanishing components of
$\tilde H$ are $\tilde H_{0r}$, 
$\tilde H_{0\hat i}$ and $\tilde H_{r\hat i}$. To the relevant order,
these components take the values:
\bear
\tilde H_{0r}&=&-i\,\sqrt{{R\over r}}\,\cot\bar\theta\,+\,
\,{i\over (\sin\bar \theta)^2}\,
\,\sqrt{{R\over r}}\,\Big(\,3\xi\,-\,{f\over \sin\bar\theta}
\,\Big)\,
-3i\,\,{\cos\bar\theta \over (\sin\bar\theta)^3}\,\,
\sqrt{{R\over r}}\,\Big(\,2\xi^2\,-\,\xi\,{f\over \sin\bar\theta}
\,\Big)\,\,+\rc\rc
&&\,+\,{i\over 2}\,R^2\,\cot\bar\theta\Bigg[\,\sqrt{{R^3\over r^3}}\,
(\partial_t\xi)^2\,-\,\sqrt{{r^3\over R^3}}\,
 (\partial_r\xi)^2\,\Bigg]\,+\,{i\over 2}\, 
\sqrt{{R\over r}}{\cos\bar\theta \over
(\sin\bar\theta)^3}\,  (\partial_{\hat i}\xi)^2\,\,, \rc\rc
\tilde H_{0\hat i}&=&{i\over 2R\sin\bar\theta}\, \,
\sqrt{{r^3\over R^3}}\,\,\hat g_{\hat i\,\hat j}\,\,
{\epsilon^{\hat j\,\hat l\,\hat m}\,\over \sqrt{\hat g}}\,\,
H_ {r\,\hat l\,\hat m}\,\,,\rc\rc
\tilde H_{r\hat i}&=&{i\over 2R\sin\bar\theta}\, 
\sqrt{{R^3\over r^3}}\,\,\hat g_{\hat i\,\hat j}\,\,
{\epsilon^{\hat j\,\hat l\,\hat m}\,\over \sqrt{\hat g}}\,\,
H_ {0\,\hat l\,\hat m}\,\,,
\label{cien}
\eear            
with $\bar\theta\equiv \bar\theta_{4,n}$. Using these results we can
compute the lagrangian for the fluctuations. After some calculation
one arrives at:
\bear
{\cal L}\,&=&\,-\,\sqrt{\hat g}\,R^3\,T_{M5}\,\cos\bar\theta\,f\,+\,
\sqrt{\hat g}\,R^3\,T_{M5}\,\big(\,\sin\bar\theta\,\big)^2
\,\,\times\rc\rc
&&\times\,{1\over 2}\,\, \Bigg[\,R^3r^{-1}(\partial_t\xi)^2\,-\,
r^2\,(\partial_r\xi)^2\,-\,(\partial_{\hat i}\xi)^2\,
+\,{1\over 2R^3r(\sin\bar\theta)^2}\,(H_{0\hat j\,\hat k})^2\,-\,\rc\rc
&&-{r^2\over 2R^6(\sin\bar\theta)^2}\,(H_{r\hat j\,\hat k})^2\,-\,
6\xi^2\,-\,{f^2\over (\sin\bar\theta)^2}\,
+\,6\,{f\xi\over \sin\bar\theta}\,\Bigg]\,\,,\rc
\label{ctuno}
\eear
where $(H_{0\hat j\,\hat k})^2$ and $(H_{r\hat j\,\hat k})^2$ are
contractions with the metric of the $S^3$. In eq. (\ref{ctuno}) we have
kept terms up to second order and we have dropped the zero-order term. 

The analysis of the equations of motion derived from eq.
(\ref{ctuno}) is similar to the one performed in section 2.3. For
this reason we will skip the details and will give directly the final
result. Let us expand $f$ and $\xi$ is spherical harmonics of $S^3$ as
in eq. (\ref{cinueve}) and let  $f_{l,m}(t,r)$ and $\zeta_{l,m}(t,r)$
denote their modes respectively. The equations of motion of these
modes can be written as:
\beq
\Bigg(R^3r^{-1}\partial_0^2\,-\,\partial_r\,r^2\,\partial_r\,
+\,{\cal M}_4\,\Bigg)\,
\pmatrix{\zeta_{l,m}\cr\cr
         {f_{l,m}\over \sin\bar\theta}}\,=\,0\,\,,
\label{ctdos}
\eeq
where the mass matrix ${\cal M}_4$ is the same as
that corresponding the  D4-D4 system (\ie\ the one of eq. 
(\ref{setenta}) for $p=4$). Notice that the wave operator on the
left-hand side of eq. (\ref{ctdos}) is formally the same as 
${\cal O}_4$ in eq. (\ref{ssiete}) (although the radius $R$ is not the
same quantity in both cases). Thus, the eigenvalues of the mass matrix
are non-negative and, actually, the same as in the D4-D4 system.
Therefore our static M-theory configurations are indeed stable.

\setcounter{equation}{0}
\section{The D3-brane in a $(p,q)$ fivebrane background}
\medskip

We are now going to study the motion of a D3-brane probe in a
background of a stack of fivebranes which are charged under both the
NS and RR three-form fields strengths of 
type IIB supergravity. This background was obtained in ref.
\cite{LuRoy} by exploiting the S-duality of type IIB supergravity and is
characterized by two coprime integers $p$ and $q$, and we will refer to it
as the  $(p,q)$ fivebrane background. It can be regarded as the one
created by an object which is a bound state of $p$ NS5-branes and $q$
D5-branes. In particular, for $(p,q)=(1,0)$ the corresponding NS5-D3
system is the analogue of the one studied in
ref. \cite{Bachas} in the type IIB theory . If, on the other hand, we take
$(p,q)=(0,1)$ we recover the D5-D3 problem studied in section 2. 

In order to describe the background, following ref. \cite{LuRoy}, let us
introduce some notations. First of all we define the quantity:
\beq
\mu_{(p,q)}\,=\,p^2\,+\,(\,q\,-\,p\chi_{0}\,)^2\,g_s^2\,,
\label{cttres}
\eeq
where $\chi_{0}$ is the asymptotic value of the RR scalar. The
``radius" $R_{(p,q)}$ for a stack of $N$ $(p,q)$ fivebranes is defined
in terms of $\mu_{(p,q)}$ as:
\beq
R^2_{(p,q)}\,=\,N\,\Bigl[\,\mu_{(p,q)}\,\Bigr]^{{1\over 2}}
\,\alpha'\,\,.
\label{ctcuatro}
\eeq
We will use $R_{(p,q)}$ to define the near-horizon harmonic function:
\beq
H_{(p,q)}(r)\,=\,{R^2_{(p,q)}\over r^2}\,\,.
\label{ctcinco}
\eeq
The near-horizon metric, in the string frame, for the stack of 
$(p,q)$ fivebranes can be written as:
\beq
ds^2\,=\,\Big[\,h_{(p,q)}(r)\,\Big]^{-{1\over 2}}\,\,
\Bigg[\,\,
\Bigl[\,H_{(p,q)}(r)\,\Bigr]^{-{1\over 4}}\,\,
(\,-dt^2\,+\,dx_{\parallel}^2\,)\,+\,
\Bigl[\,H_{(p,q)}(r)\,\Bigr]^{{3\over 4}}\,\,
(\,dr^2\,+\,r^2\,d\Omega_{3}^2\,)\,\Bigg]\,\,,
\label{ctseis}
\eeq
where the function $h_{(p,q)}(r)$ is given by:
\beq
h_{(p,q)}(r)\,=\,{\mu_{(p,q)}\over
p^2\,\,\Bigl[\,H_{(p,q)}(r)\,\Bigr]^{{1\over 2}}
\,+\,(\,q\,-\,p\chi_{0}\,)^2\,g_s^2\,\,
\Bigl[\,H_{(p,q)}(r)\,\Bigr]^{-{1\over 2}}}\,\,.
\label{ctsiete}
\eeq
To simplify the equations that follow we shall take 
from now on  $g_s=1$ and $\chi_0\,=\,0$. (The dependence on $g_s$ and 
$\chi_0$ can be easily restored). Other fields of this background
include the dilaton:
\beq
e^{-\phi}\,=\,h_{(p,q)}(r)\,\,,
\label{ctocho}
\eeq
and the RR scalar:
\beq
\chi\,=\,{pq\over \mu_{(p,q)}}\,
\Big(\,\Bigl[\,H_{(p,q)}(r)\,\bigr]^{{1\over 2}}\,-\,
\bigl[\,H_{(p,q)}(r)\,\bigr]^{-{1\over 2}}
\,\Big)\,h_{(p,q)}(r)\,\,.
\label{ctnueve}
\eeq
In addition we have non-zero NS and RR three-form field strengths. Let
us call $B$ and $C^{(2)}$ to their two-form potentials respectively.
If we take coordinates on the three-sphere as in eq.  (\ref{cinco}),
this potentials can be taken as:
\beq
B\,=\,-pN\alpha'\,C_5(\theta)\,\epsilon_{(2)}\,\,,
\,\,\,\,\,\,\,\,\,\,\,\,\,\,\,\,\,\,\,\,\,\,\,\,\,
C^{(2)}\,=\,-qN\alpha'\,C_5(\theta)\,\epsilon_{(2)}\,\,,
\label{ctdiez}
\eeq
where $C_5(\theta)$ is the function defined in eqs. (\ref{seis}) and
(\ref{siete}), \ie\ 
$C_5(\theta)\,=\,\sin \theta\cos\theta\,-\,\theta$. 

The action of a D3-brane probe in the above background is the sum of the
Dirac-Born-Infeld and  Wess-Zumino terms. The latter now includes the
coupling of the brane to the RR potential $C^{(2)}$ and to the RR scalar
$\chi$:
\beq
S\,=\,-T_3\,\int d^4\sigma\,e^{-\phi}\,\sqrt{-{\rm det}(g+{\cal
F})}\,+\, T_3\,\int \Big[\,{\cal F}\wedge C^{(2)}\,
+\,{1\over 2}\chi{\cal F}\wedge{\cal F}\,\Big]\,\,,
\label{ctonce}
\eeq
with ${\cal F}$ being:
\beq
{\cal F}\,=\,dA\,-\,B\,=\,F\,-\,B\,\,.
\label{ctdoce}
\eeq

\medskip
\subsection{Quantization conditions}
\medskip
The analysis of the action (\ref{ctonce}) for the $(p,q)$ fivebrane
background was performed in ref. \cite{Llatas} (for the NS5-D3 system
see refs. \cite{Joan,Pelc}). Here we shall choose our
worldvolume coordinates as in   (\ref{doce}) and we will look for
solutions of the equations of motion with constant $\theta$. Notice that,
as now our background contains non-zero NS and RR forms, it is natural to
expect that the worldvolume gauge field $F$ has both electric and magnetic
components. The latter can be determined by means of the flux
quantization condition   (\ref{uno}), whereas the electric wordlvolume
field is constrained by the condition (\ref{dsiete}). Accordingly we
shall first require that:
\beq
\int_{S^2}\,F\,=\,{2\pi n_1\over T_f}\,\,,
\label{cttrece}
\eeq
with $n_1\in\ZZ$. It is rather simple to solve this condition. We only
have to take $F$ as:
\beq
F\,=\,\pi n_1\alpha'\epsilon_{(2)}\,+\,F_{0,r}dt\wedge dr\,\,,
\label{ctcatorce}
\eeq
where we have assumed that the electric worldvolume field has only
components along the radial direction. By using the definition of 
${\cal F}$ in eq. (\ref{ctdoce}) and the expression of the $B$ field
in eq. (\ref{ctdiez}), one easily verifies that eq. (\ref{ctcatorce})
is equivalent to the following expression for ${\cal F}$:
\beq
{\cal F}\,=\,f_{12}(\theta)\epsilon_{(2)}\,+\,F_{0,r}dt\wedge dr\,\,,
\label{ctquince}
\eeq
with $f_{12}(\theta)$ being:
\beq
f_{12}(\theta)\,\equiv\,pN\alpha'C_5(\theta)\,+\,\pi n_1\alpha'\,\,.
\label{ctdseis}
\eeq
As in our previous examples, let us  assume that the angle $\theta$
depends only on the radial coordinate $r$. By substituting the ansatz 
(\ref{ctquince}) in eq. (\ref{ctonce}), one can find the form of the 
lagrangian density:
\bear
{\cal L}(\theta, F)\,&=&\,-T_3\,\,\sqrt{\hat g}\,\Bigg[\,
\sqrt{r^4\,\Bigl[\,H_{(p,q)}(r)\,\Bigr]^{{3\over 2}}\,(\sin\theta)^4
\,+\,e^{-\phi}\,f_{12}(\theta)^2}\,\times\rc\rc
&&\times\,\sqrt{\Bigl[\,H_{(p,q)}(r)\,\Bigr]^{{1\over 2}}
(1+r^2\theta'^2)\,-\,e^{-\phi}\,F_{0,r}^2}\,+\,\rc\rc
&&+\,(qN\alpha'C_5(\theta)\,-\,\chi f_{12}(\theta))\,F_{0,r}
\,\Bigg]\,\,,
\label{ctdsiete}
\eear
where $\hat g$ is the determinant of the metric of a unit $S^{2}$.
Next, we make use of the electric quantization condition 
(\ref{dsiete}) and require that:
\beq
\int_{S^2}\,
d^2\,\theta\,\,\,
{\partial{\cal L}\over \partial F_{0,r}}\,=\,
n_2\,T_f\,\,,
\label{ctdocho}
\eeq
where $n_2$ is another integer. By plugging the lagrangian density 
(\ref{ctdsiete}) into (\ref{ctdocho}),  one can obtain $F_{0,r}$ as a
function of $\theta(r)$ and of the integers $n_1$ and $n_2$. Actually,
one can eliminate in this way $F_{0,r}$ from the expression of the
hamiltonian $H$, which can be obtained from ${\cal L}$ by means of a
Legendre transformation (see eq. (\ref{vtres})). The resulting
hamiltonian can be put in the form:
\bear
&&H\,=\,T_3\Omega_2\,\int dr\sqrt{1\,+\,r^2\theta'^2}\,\times\,\rc\rc
&&\times\,\sqrt{R^4_{(p,q)}\,(\sin\theta)^4\,+\,
[\mu_{(p,q)}]^{-1}\big[\,\big(pf_{12}(\theta)+q\Pi(\theta)\big)^2+
H_{(p,q)}(r)\big(qf_{12}(\theta)-p\Pi(\theta)\big)^2\,\big]}\,\,,
\rc\rc
\label{ctdnueve}
\eear
where $\Pi(\theta)$ is the function:
\beq
\Pi(\theta)\,\equiv\,qN\alpha'C_5(\theta)\,+\,\pi n_2\alpha'\,\,.
\label{ctveinte}
\eeq
The solutions of the equations of motion with $\theta={\rm constant}$
can be obtained by solving the equation 
$\partial H/ \partial\theta=0$ for  $\theta'=0$. A glance at the
right-hand side of eq. (\ref{ctdnueve}) reveals immediately that these
solutions only exist if the $r$-dependent term inside the square root 
in (\ref{ctdnueve})  is zero. We thus get the condition:
\beq
qf_{12}(\theta)\,=\,p\Pi(\theta)\,\,,
\label{ctvuno}
\eeq
which, after using eqs. (\ref{ctdseis}) and (\ref{ctveinte}), is
equivalent to the following relation between the integers $n_1$ and
$n_2$:
\beq
qn_1\,=\,pn_2\,\,.
\label{ctvdos}
\eeq
But,  as $p$ and $q$ are coprime integers, the only possibility to
fulfill eq. (\ref{ctvdos}) is that $n_1$ and $n_2$ be of the form:
\beq
n_1\,=\,p\,n\,,
\,\,\,\,\,\,\,\,\,\,\,\,\,\,\,\,\,\,\,
n_2\,=\,qn\,\,,
\label{ctvtres}
\eeq
with $n\in\ZZ$. Thus our two quantization integers $n_1$ and $n_2$ are
not independent and they can be put in terms of another integer $n$.
By using the relations (\ref{ctvtres}), one can rewrite 
$f_{12}(\theta)$ and $\Pi(\theta)$ in terms of $n$:
\beq
f_{12}(\theta)\,=\,pN\alpha'\,{\cal C}_{5,n}(\theta)\,,
\,\,\,\,\,\,\,\,\,\,\,\,\,\,\,\,\,\,\,
\Pi(\theta)\,=\,qN\alpha'\,{\cal C}_{5,n}(\theta)\,\,,
\label{ctvcuatro}
\eeq
where ${\cal C}_{5,n}(\theta)$ is the function defined in eq. 
(\ref{vuno}). If we now substitute these expressions into the
hamiltonian (\ref{ctdnueve}), we get the following expression of $H$:
\beq
H\,=\,T_3\,\Omega_2\,R_{(p,q)}^2\,\int dr\,
\sqrt{1+r^2\theta'^2}\,
\sqrt{(\sin\theta)^4\,+\,
\Big(\,{\cal C}_{5,n}(\theta)\Big)^2}\,\,.
\label{ctvcinco}
\eeq
Apart from a global coefficient, this hamiltonian is the same as the
one in eq. (\ref{vcuatro}) for the D5-brane background. Thus, the
energy is clearly minimized for $\theta\,=\,\bar\theta_{5,n}$, where
the $\bar\theta_{5,n}$'s are the angles written in eq. 
(\ref{tocho}) . The energy densities for these angles are easily
computed from eq. (\ref{ctvcinco}). One gets:
\beq
{\cal E}_{5,n}^{(p,q)}\,=\,{NT_{(q,p)}\over \pi}\,
\sin\Big[\,{n\over N}\,\pi\Big]\,\,,
\label{ctvseis}
\eeq
where $T_{(q,p)}$ is the tension of the $(q,p)$-string which, for
arbitrary values of $g_s$ and $\chi_0$, is given by:
\beq
T_{(q,p)}\,=\,\sqrt{(q-p\chi_0)^2\,+\,{p^2\over g_s^2}}\,\,\,
T_f\,\,.
\label{ctvsiete}
\eeq
By comparing eqs. (\ref{tnueve}) and (\ref{ctvseis}) it follows that 
${\cal E}_{5,n}^{(p,q)}$ can be obtained from ${\cal E}_{5,n}$ by
substituting $T_f$ by $T_{(q,p)}$. In particular, if
$N\rightarrow\infty$ the energy density ${\cal E}_{5,n}^{(p,q)}$ equals
$nT_{(q,p)}$ and, thus, the configurations we have found can be
regarded as  bound states of $n$ $(q,p)$-strings. It is also easy to
get the worldvolume electric field $\bar F_{0,r}$ of our solutions. It
takes the form:
\beq
\bar F_{0,r}\,=\,{(q-p\chi_0)g_s\over 
\sqrt{p^2+(q-p\chi_0)^2g_s^2}}\,\,
\cos\,\Big[\,{n\over N}\,\pi\Big]\,\,.
\label{ctvocho}
\eeq
It is also clear from the expression of the hamiltonian in 
(\ref{ctvcinco}) that one can represent it as in eq. (\ref{csiete})
and, as a consequence, one can find a bound for the energy whose
saturation gives rise to a BPS condition. As the only difference
between  the hamiltonian (\ref{ctvcinco}) and that of eq. 
(\ref{vcuatro}) for the D5-brane is a global coefficient, it follows
that  the BPS differential equation is just the one displayed in eq. 
(\ref{ciuno}) for $p=5$. Its solution is given in eq.
(\ref{citres}) and, again,  includes our wrapped configurations as 
particular cases. Moreover, these $\theta\,=\,\bar\theta_{p,n}$
configurations can be regarded as the $r\rightarrow 0$ limit of the
general solution. Actually, in ref. \cite{Llatas} the condition 
(\ref{ctvuno})  and the form (\ref{ctvcinco}) of the hamiltonian were
obtained by using the S-duality of the worldvolume action. It was also
checked in this reference that the D3-brane configurations which
saturate the bound preserve $1/4$ of the bulk supersymmetry.

\medskip
\subsection{Stability}
\medskip

The static configurations of the D3-brane studied above are stable
under small perturbations, as one can check following the same steps
as in sections 2.3 and 3.2. First of all, we parametrize the angle
fluctuations as:
\beq
\theta\,=\,\bar\theta_{5,n}\,+\,\xi\,\,,
\label{ctvnueve}
\eeq
whereas the gauge field fluctuates as:
\beq
{\cal F}\,=\,\big[\,f_{12}(\theta)\,+\,g\,\big]\,\epsilon_{(2)}\,+\,
\big[\,\bar F_{0,r}\,+\,f\,\big]\,dt\wedge dr\,\,.
\label{cttreinta}
\eeq
The angle fluctuation $\xi$ and the electric (magnetic) field
fluctuation $f$ ($g$) are supposed to be small and only terms up to
second order are retained in the lagrangian. The corresponding
equations of motion involve now the wave operator 
${\cal O}_5^{(p,q)}$, which acts on any function $\psi$  as:
\beq
{\cal O}_5^{(p,q)}\,\psi\,\equiv\,R^2_{(p,q)}\,\partial_0^2\,\psi\,-\,
\partial_r(r^2\partial_r\psi)\,\,.
\label{cttuno}
\eeq 
Notice that ${\cal O}_5^{(p,q)}$ is obtained from ${\cal O}_5$ in eq. 
(\ref{ssiete}) by means of the substitution 
$R\rightarrow R_{(p,q)}$. Let us combine $\xi$, $f$ and $g$ into the
field $\eta$, defined as:
\bear
\eta&\equiv&{1\over p^2\,H_{(p,q)}(r)\,\,+\,q^2\sin^2\bar\theta_{5,n}}
\,\,\times\rc\rc
&&\times\Big[\,[\mu_{(p,q)}]^{{1\over 2}}\,
\big(\,q\sin\bar\theta_{5,n}\,f\,-\,{p\over r^2}\,g\,\big)\,+\,
2q^2\sin^2\bar\theta_{5,n}\,\xi\,\Big]\,\,,
\label{cttdos}
\eear
and let us expand $\xi$ and $\eta$ is spherical harmonics of $S^2$. If 
$\zeta_{l,m}$ and $\eta_{l,m}$ denote their modes respectively, one
can prove after some calculation that the equations of motion for 
$\zeta_{l,m}$ and $\eta_{l,m}$ can be written as:
\beq
\Big(\,{\cal O}_5^{(p,q)}\,+\,{\cal M}_5\,\Big)\,
\pmatrix{\zeta_{l,m}\cr\eta_{l,m}}\,=\,0\,\,. 
\label{ctttres}
\eeq
In eq. (\ref{ctttres}) ${\cal M}_5$ is the matrix defined in eq. 
(\ref{setenta}), whose eigenvalues, as proved in section 2.3, are always
non-negative. There is also a decoupled mode $\sigma$, whose
expression in terms of $\xi$, $f$ and $g$ is:
\bear
\sigma&\equiv&
{\sin\bar\theta_{5,n}
\over r[\,
p^2\,H_{(p,q)}(r)\,\,+\,q^2\sin^2\bar\theta_{5,n}\,]}
\,\,\times\rc\rc
&&\times\Big[\,[\mu_{(p,q)}]^{{1\over 2}}\,
\big(\,q\sin\bar\theta_{5,n}\,g\,-\,p\,R^2_{(p,q)}\,f\,\big)\,-\,
2pq\,R^2_{(p,q)}\sin\bar\theta_{5,n}\,\xi\,\Big]\,\,.
\label{cttcuatro}
\eear
The equation of motion of $\sigma$ can be written as:
\beq
\Big(\,{\cal O}_5^{(p,q)}\,+\,l(l+1)\,\Big)\,\sigma_{l,m}\,=\,0\,\,,
\label{cttcinco}
\eeq
where $\sigma_{l,m}$ are the modes of the expansion of $\sigma$ in
$S^2$-spherical harmonics. It is evident from eq. (\ref{cttcinco})
that the mass eigenvalues of  $\sigma_{l,m}$ are non-negative, which
confirms that the configurations around which we are expanding are
stable.

\medskip
\section{Summary and discussion}
\medskip

In this paper we have studied certain  configurations of branes  which
are partially wrapped on spheres. These spheres are placed on the
transverse region of some supergravity background, and their positions,
characterized by a polar angle which measures their latitude in a system
of spherical coordinates, are quantized and given by a very specific set
of values. We have checked that our configurations are stable by analyzing
their behaviour under small fluctuations and, by studying their energy, we
concluded that they can be regarded as a bound state of strings or, in
the case of the M5-M5 system, M2-branes. We have verified this fact
explicitly in appendix B for the case of a wrapped D3-brane in the
background of a NS5-brane. Indeed, we have proved that, by embedding a
D1-brane in a fuzzy two-sphere in the NS5-brane background, one obtains
exactly the same energies and allowed polar angles as for a wrapped
D3-brane in the same geometry. Clearly, a similar description of all the
cases studied here would be desirable and would help to understand
more precisely the r\^ole of noncommutative geometry in the formation of
these bound states. In this sense it is interesting to point out that
the polarization of multiple fundamental strings in a RR background was
studied in ref. \cite{Schiappa}

Contrary to ref. \cite{Bachas}, the problems treated here do not have a
CFT description  to compare with. Thus, we do not know to what
extent we can trust our Born-Infeld results. However, one could argue
that we have followed the same methodology as in ref. \cite{Bachas} and,
actually, our configurations can be connected to the ones in
\cite{Bachas} by string dualities. Moreover, the BPS nature of our
configurations make us reasonably confident of the correctness of our
conclusions. 

The presence of a non-trivial supergravity background is of crucial
importance in our analysis. Indeed, these backgrounds induce worldvolume
gauge fields on the brane probes, which prevent their collapse. The
stabilization mechanisms found here are a generalization of the one
described in refs. \cite{Bachas, Pavel}, and are based on a series of
quantization rules which determine the values of the worldvolume gauge
fields. We have reasons to believe that our results are generic and can
be extended to other geometries such as, for example, the ones generated
by the Dp-D(p-2) bound states \cite{MR}. Another interesting question is
the implications of our results in a holographic description of gauge
theories. In our opinion the study of these topics could  enrich
our knowledge of the brane interaction dynamics.

\section{ Acknowledgments}
We are grateful to J. L. Barbon, C. Gomez, J. Mas, T. Ortin and J. M.
Sanchez de Santos for discussions.   This work was
supported in part by DGICYT under grant PB96-0960,  by CICYT under
grant  AEN99-0589-CO2 and by Xunta de Galicia under  grant
PGIDT00-PXI-20609.

\vskip 2cm                                               
{\Large{\bf APPENDIX A}}                                 
\vskip .5cm                                              
\renewcommand{\theequation}{\rm{A}.\arabic{equation}}  
\setcounter{equation}{0}  
In this appendix we collect the expressions of the functions 
$\Lambda_{p,n}(\theta)$ for $0\le p\le 5$. They are:
\bear
\Lambda_{0,n}(\theta)&=&-{2\over 5}\,\Big[\,\cos\theta\,\Big(\,
3\sin^4\theta\,+\,4\sin^2\theta\,+\,8\,\Big)\,+\,8\,\Big(\,
2\,{n\over N}\,-\,1\Big)\,\Big]\,\,,\rc\rc
\Lambda_{1,n}(\theta)&=&-{5\over 4}\,\Big[\,\cos\theta\,\Big(\,
\sin^3\theta\,+\,{3\over 2}\,\sin\theta\,\Big)\,+\,
{3\over 2}\,\Big(\,{n\over N}\,\pi\,-\,\theta\,\Big)\,\Big]\,\,, \rc\rc
\Lambda_{2,n}(\theta)&=&-{4\over 3}\,\Big[\,\cos\theta\,\Big(\,
\sin^2\theta\,+\,2\,\Big)\,+\,2\,\Big(\,
2\,{n\over N}\,-\,1\Big)\,\Big]\,\,,\rc\rc
\Lambda_{3,n}(\theta)&=&-{3\over 2}\,\Big[\,\cos\theta\,\sin\theta\,+\,
{n\over N}\,\pi\,-\,\theta\,\Big]\,\,,\rc\rc
\Lambda_{4,n}(\theta)&=&- 2\,\Big[\,\cos\theta\,+\,
2\,{n\over N}\,-\,1\Big]\,\,,\rc\rc
\Lambda_{5,n}(\theta)&=&\theta\,-\,{n\over N}\,\pi\,\,.
\label{apauno}
\eear
The functions ${\cal C}_{p,n}(\theta)$ and $ C_p(\theta)$ can be
easily obtained from (\ref{apauno}) by using their relation with the 
$\Lambda_{p,n}(\theta)$'s (see eqs. (\ref{vseis}) and 
(\ref{vuno})).

\vskip 1cm                                               
{\Large{\bf APPENDIX B}}                                 
\vskip .5cm                                              
\renewcommand{\theequation}{\rm{B}.\arabic{equation}}  
\setcounter{equation}{0}  
In this appendix we will show how one can represent the wrapped branes
studied in the main text as a bound state of strings. We will make use
of the Myers polarization mechanism \cite{Myers}, in which the strings
are embedded in a noncommutative space. Actually, we will only consider a
particular case of those analyzed in sects. 2-4, namely the one of
section 4 with $p=1$, $q=\chi_0=0$, \ie\ the D3-brane in the
background of the NS5-brane. For convenience we will choose a new set
of coordinates to parametrize the space transverse to the NS5. Instead
of using the radial coordinate $r$ and the three angles $\theta^1$, 
 $\theta^2$ and  $\theta$ (see eq. (\ref{cinco})), we will work with
four cartesian coordinates $z,x^1, x^2, x^3$, which, in terms of the
spherical coordinates, are given by: 
\bear
z&=&r\cos\theta\,\,,\cr
x^1&=&r\sin\theta\cos\theta^2\,\,,\cr
x^2&=&r\sin\theta\sin\theta^2\cos\theta^1\,\,,\cr
x^2&=&r\sin\theta\sin\theta^2\sin\theta^1\,\,.
\label{apbuno}
\eear
Conversely, $r$ and $\theta$ can be put in terms of the new
coordinates as follows:
\bear
r\,&=&\,
\sqrt{(z)^2\,+\,(x^1)^2\,\,+(x^2)^2\,\,+(x^3)^2\,\,}\,\,,\cr\cr\cr
\tan\theta&=&{\sqrt{(x^1)^2\,\,+(x^2)^2\,\,+(x^3)^2\,\,}\over z}\,\,.
\label{apbdos}
\eear
In what follows some of our expressions will contain $r$ and $\theta$.
It should be understood that they are given by the functions of 
$(\,z,x^i\,)$ written in eq. (\ref{apbdos}). The near-horizon metric
and the dilaton for a stack of N NS5-branes are (see eqs. 
(\ref{ctseis}) and (\ref{ctocho})):
\bear
ds^2&=&-dt^2\,+\,dx_{\parallel}^2\,+\,
{N\alpha'\over r^2}\,\,
\Big(\,(dz)^2\,+\,(dx^1)^2\,+\,(dx^2)^2\,+\,(dx^3)^2\,\Big)\,\,,\rc\rc
e^{-\phi}\,&=&\,{r\over \sqrt{N\alpha'}}\,\,.
\label{apbtres}
\eear
Moreover, the non-vanishing components of the $B$ field in the new 
coordinates can be obtained from eq. (\ref{ctdiez}). They are: 
\beq
B_{x^ix^j}\,=\,N\alpha'\,{C_5(\theta)\over r^3\sin^3\theta}\,\,
\epsilon_{ijk}\,x^k\,\,.
\label{apbcuatro}
\eeq

According to our analysis of section 4, the wrapped D3-brane in this
background can be described as a bound state of D1-branes. Thus it is
clear that we must consider a system of $n$ D1-branes, moving in the
space transverse to the stack of $N$ Neveu-Schwarz fivebranes. We will
employ a static gauge where the two worldsheet coordinates will be
identified with $t$ and $z$. The Myers proposal for the action of this
system  is:
\bear
S_{D1}\,&=&\,-T_1\,\int dtdz\,{\rm STr}\,\Bigg[\,
e^{-\phi}\,\sqrt{-{\rm det}\Big(
P\big[E_{ab}+E_{ai}\,(\,Q^{-1}-\delta\,)^{ij}\,E_{jb}\big]
+\lambda F_{ab}\Big)\,{\rm det}\Big(Q^i_{\,\,j}\Big)\,}\,\,
\Bigg]\,\,,\rc\rc
\label{apbcinco}
\eear
where we are adopting the conventions of ref. \cite{Myers}. In eq.
(\ref{apbcinco}) $\lambda\,=\,2\pi\alpha'\,=\,1/T_f$, $F_{ab}$ is the
worldsheet gauge field strength (which we will assume that is zero in our
case),  $P$ denotes the pullback of the spacetime tensors to the D1-brane
worldsheet and STr represents the Tseytlin symmetrized trace of
matrices \cite{STr}. The indices $a,b,\cdots$ correspond to directions
parallel to the worldsheet (\ie\ to $t$ and $z$), whereas $i,j\cdots$
refer to directions transverse to the D1-brane probe. The tensor
$E_{\mu\nu}$ is defined as:
\beq
E_{\mu\nu}\,=\,G_{\mu\nu}\,+\,B_{\mu\nu}\,\,,
\label{apbseis}
\eeq
where $G_{\mu\nu}$ is the background metric. Let $\phi^i$ denote the
transverse scalar fields, which are matrices taking values in the
adjoint representation of $U(n)$. Then $Q^i_{\,\,j}$ is defined as:
\beq
Q^i_{\,\,j}=\,\delta^i_{\,\,j}\,+\,i\lambda\,[\,\phi^i\,,\,\phi^k\,]
\,E_{kj}\,\,. 
\label{apbsiete}
\eeq
As in ref. \cite{Myers}, transverse indices are raised with $E^{ij}$,
where 
$E^{ij}$ denotes the inverse of $E_{ij}$, \ie\ 
$E^{ik}E_{kj}\,=\,\delta^i_{\,j}$.

Let us now make the standard identification between the transverse
coordinates $x^i$ and the scalar fields $\phi^i$, namely:
\beq
x^i\,=\,\lambda\,\phi^i\,\,.
\label{apbocho}
\eeq
Notice that, after the identification (\ref{apbocho}), the $x^i$'s
become noncommutative coordinates represented by matrices. Actually,
as in ref. \cite{Myers}, we will make the following ansatz for the scalar
fields:
\beq
\phi^i\,=\,{f\over 2}\,\alpha^i\,\,,
\label{apbnueve}
\eeq
where $f$ is a c-number to be determined and
the $\alpha^i$'s are $n\times n$ matrices corresponding to the
$n$-dimensional irreducible representation of $su(2)$:
\beq
[\,\alpha^i\,,\,\alpha^j\,]\,=\,2i\epsilon_{ijk}\,\alpha^k\,\,.
\label{apbdiez}
\eeq
As the quadratic Casimir of the $n$-dimensional irreducible
representation of $su(2)$ is $n^2-1$, we can write:
\beq
(\alpha^1)^2\,\,+(\alpha^2)^2\,\,+(\alpha^3)^2\,=\,
(n^2\,-\,1)\,I_n\,\,,
\label{apbonce}
\eeq
where $I_n$ is the $n\times n$ unit matrix. By using eqs.
(\ref{apbocho}) and (\ref{apbnueve}) in (\ref{apbonce}), we get:
\beq
(x^1)^2\,\,+(x^2)^2\,\,+(x^3)^2\,=\,{\lambda^2\,f^2\over 4}\,
(n^2\,-\,1)\,I_n\,\,,
\label{apbdoce}
\eeq
which shows that, with our ansatz, the $x^i$'s are coordinates of a
fuzzy two-sphere of radius $\lambda\,f\,\sqrt{n^2-1}/2$. On the other
hand, if we treat the $x^i$'s as commutative coordinates, it is easy
to conclude from eqs. (\ref{apbuno}) and (\ref{apbdos}) that the
left-hand side of (\ref{apbdoce}) is just $(r\sin\theta)^2$. In view of
this, when the  $x^i$'s  are non-commutative  we should identify the
expression written in eq. (\ref{apbdoce}) with 
$(r\sin\theta)^2\,I_n$. Thus, we put:
\beq
{f\over 2}\,=\,{r\sin\theta\over \lambda\sqrt{n^2\,-\,1}}\,\,.
\label{apbtrece}
\eeq
Notice that, as can be immediately inferred from eq. (\ref{apbdos}),
$r$ and $\theta$ depend on the $x^i$'s through the sum
$\sum_i\,(x^i)^2$, which is proportional to the $su(2)$ quadratic
Casimir. Then, as matrices, $r$ and $\theta$ are  multiple of  the
unit matrix and, thus, we can consider them as commutative
coordinates. This, in particular,  means that the  elements of the
metric tensor $G_{\mu\nu}$ are also commutative, whereas, on the
contrary, the components of the $B$ field have a non-trivial matrix
structure. By substituting our ansatz in eqs. (\ref{apbtres}) and 
(\ref{apbcuatro}), we get the following expression for the transverse
components of the $E_{\mu\nu}$ tensor:
\beq
E_{ij}\,=\,{N\alpha'\over r^2}\,\Big[\,
\delta^i_{\,\,j}\,\,+\,\,{1\over\sqrt{n^2\,-\,1}}\,\,\,
{C_5(\theta)\over \sin^2\theta}\,\,
\epsilon_{ijk}\,\alpha^k\,\Big]\,\,.
\label{apbcatorce}
\eeq
The quantities $Q^i_{\,\,j}$, defined in eq. (\ref{apbsiete}), can be
readily obtained from eq. (\ref{apbcatorce}), namely: 
\beq
Q^i_{\,\,j}\,=\,\Big(\,1\,+\,{N\over \pi}\,
{C_5(\theta)\over \sqrt{n^2\,-\,1}}\,\Big)\,\delta^i_{\,\,j}\,-\,
{N\over \pi}\,{C_5(\theta)\over (n^2\,-\,1)^{3/2}}
\,\alpha^j\,\alpha^i\,-\,{N\over \pi}\,
{\sin^2\theta\over n^2-1}\,\,\epsilon_{ijk}\,\alpha^k\,\,.
\label{apbquince}
\eeq
In order to compute the pullback appearing in the first determinant of
the right-hand side of eq. (\ref{apbcinco}), we need to characterize
the precise embedding of the D1-brane in the transverse
non-commutative space. Actually, it is straightforward to write our
ansatz for the $x^i$'s as:
\beq
x^i\,=\,z\,\,{\tan\theta\over  \sqrt{n^2\,-\,1}}\,\,\,\alpha^i\,\,.
\label{apbdseis}
\eeq
Moreover, the kind of configurations we are looking for have constant
$\theta$ angle. Thus, eq.  (\ref{apbdseis}) shows that, in this case,
the $x^i$'s are linear functions of the worldsheet coordinate $z$. By
using this result it is immediate to find the expression of the first
determinant in (\ref{apbcinco}). One gets:
\beq
-{\rm det}\Big(
P\big[E_{ab}+E_{ai}\,(\,Q^{-1}-\delta\,)^{ij}\,E_{jb}\big]\,
\Big)\,=\,{N\alpha'\over r^2}\,+\,
{\tan^2\theta\over n^2-1}\,\,\,
\alpha^i\,\Big[Q^{-1}\Big]_{ij}\,\alpha^j\,\,,
\label{apbdsiete}
\eeq
where $Q^{-1}$ satisfies 
$Q^{ij}\Big[Q^{-1}\Big]_{jk}\,=\,\delta^i_{\,\,k}$ with $Q^{ij}$ 
being:
\beq
Q^{ij}\,=\,E^{ij}\,+\,i\lambda\,\,[\,\phi^i\,,\,\phi^j\,]\,\,.
\label{apbdocho}
\eeq
As expected on general grounds, a system of D-strings can model a
D3-brane only when the number $n$ of D-strings is very large. Thus, if we
want to make contact with our results of section 4, we should consider
the limit in which $n\rightarrow\infty$  and keep
only the leading terms in the $1/n$ expansion. Therefore, it this
clear that, in this limit, we can replace $n^2-1$ by $n^2$ in all 
our previous expressions. Moreover, as argued in ref. \cite{Myers}, the
leading term in a symmetrized trace of $\alpha$'s of the form 
${\rm STr}\big(\,(\alpha^i\alpha^i)^m\,\big)$ is $n(n^2)^m$. Then, at
leading order in $1/n$, one can make the following replacement inside
a symmetrized trace:
\beq
\alpha^i\alpha^i\,\,\rightarrow\,\,n^2\,I_n\,\,.
\label{apbdnueve}
\eeq
With this substitution the calculation of the action (\ref{apbcinco})
drastically simplifies. So, for example, by using (\ref{apbquince}),
one can check that, in the second term under the square root of 
(\ref{apbcinco}), we should make the substitution:
\beq
{\rm det} \Big(\,Q^i_{\,\,j}\,\Big)\,\rightarrow\,
\Bigg(\,{N\over \pi n}\,\Bigg)^2\,\,
\Bigg[\,\big(\sin\theta\big)^4\,+\,
\big(\,C_5(\theta)\,+\,{\pi n\over N}\,\big)^2\,\Bigg]\,I_n\,\,.
\label{apbveinte}
\eeq
Moreover, as 
${\cal C}_{5,n}(\theta)\,=\,C_5(\theta)\,+\,{\pi n\over N}$, eq. 
(\ref{apbveinte}) is equivalent to:
\beq
{\rm det} \Big(\,Q^i_{\,\,j}\,\Big)\,\rightarrow\,
\Bigg(\,{N\over \pi n}\,\Bigg)^2\,\,
\Bigg[\,\big(\sin\theta\big)^4\,+\,
\big(\,{\cal C}_{5,n}(\theta)\,\big)^2\,\Bigg]\,I_n\,\,.
\label{apbvuno}
\eeq

We must now perform the substitution (\ref{apbdnueve}) on the
right-hand side of eq. (\ref{apbdsiete}). First of all, we must
invert the matrix of eq. (\ref{apbdocho}). Actually, it is not difficult
to obtain the expression of $E^{ij}$. After some calculation one gets:
\beq
E^{ij}\,=\,{r^2\over N\alpha'}\,\,
{\sin^4\theta\over\sin^4\theta\,+\,\Big(\,C_5(\theta)\,)^2}\,\,
\Bigg[\,\delta^i_{\,\,j}\,\,+\,\,
{\Big(\,C_5(\theta)\,)^2\over n^2\sin^4\theta}\,\alpha^i\,\alpha^j
\,\,-\,\,{C_5(\theta)\over n\sin^2\theta}\,\,
\epsilon_{ijk}\,\alpha^k\,\Bigg]\,\,.
\label{apbvdos}
\eeq
Plugging this result on the right-hand side of eq. (\ref{apbdocho}),
and adding the commutator of the scalar fields, one immediately
obtains $Q^{ij}$. By inverting this last matrix one arrives at the
following expression of $[Q^{-1}]_{ij}$:
\beq
[Q^{-1}]_{ij}\,=\,{N\alpha'\over r^2}\,\,\,
{\sin^4\theta\,+\,\Big(\,C_5(\theta)\,)^2\over 
(1+a^2)\sin^4\theta}\,\,\Bigg[\,\delta^i_{\,\,j}\,\,
+\,\,{a^2-b\over n^2(1+b)}\,\alpha^i\,\alpha^j\,\,+\,\,
{a\over n}\,\epsilon_{ijk}\,\alpha^k\,\Bigg]\,\,,
\label{apbvtres}
\eeq
where, at leading order, $a$ and $b$ are given by:
\beq
a={N\over \pi n\sin^2\theta}\,\,\Big[\,\sin^4\theta\,+\,
C_5(\theta)\,{\cal C}_{5,n}(\theta)\,\Big]\,\,,
\,\,\,\,\,\,\,\,\,\,\,\,\,\,\,\,\,
b\,=\,{\Big(\,C_5(\theta)\,)^2\over\sin^4\theta}\,\,.
\label{apbvcuatro}
\eeq
By contracting $[Q^{-1}]_{ij}$ with $\alpha^i\alpha^j$ and applying
the substitution (\ref{apbdnueve}), one gets a remarkably simple result:
\beq
\alpha^i\,\Big[Q^{-1}\Big]_{ij}\,\alpha^j\,\rightarrow\,
n^2\,{N\alpha'\over r^2}\,I_n\,\,.
\label{apbvcinco}
\eeq
By using eq. (\ref{apbvcinco}), one immediately concludes that we should
make the following substitution:
\beq
-{\rm det}\Big(
P\big[E_{ab}+E_{ai}\,(\,Q^{-1}-\delta\,)^{ij}\,E_{jb}\big]\,
\Big)\,\,\rightarrow\,
{N\alpha'\over r^2\cos^2\theta}\,I_n\,\,.
\label{apbvseis}
\eeq
It is now straightforward to find the action of the D1-branes in the
large $n$ limit. Indeed, by using eqs. (\ref{apbvuno}) and
(\ref{apbvseis}), one gets:
\beq
S_{D1}\,=\,-T_1\,\int dtdz\,{N\over \pi\cos\theta}\,\,
\sqrt{\big(\sin\theta\big)^4\,+\,
\big(\,{\cal C}_{5,n}(\theta)\,\big)^2}\,\,.
\label{apbvsiete}
\eeq
From eq. (\ref{apbvsiete}) one can immediately obtain the hamiltonian
of the D-strings. In order to compare this result with the one
corresponding to the wrapped D3-brane, let us change the worldsheet
coordinate from $z$ to $r\,=\,z/\cos\theta$. Recalling that $\theta$
is constant for the configurations under study  and using that 
$T_1/\pi\,=\,4\pi\alpha'\,T_3\,=\,T_3\Omega_2\,\alpha'$, we get the
following hamiltonian:
\beq
H\,=\,T_3\Omega_2\,N\alpha'\,\int dr\,
\sqrt{\big(\sin\theta\big)^4\,+\,
\big(\,{\cal C}_{5,n}(\theta)\,\big)^2}\,\,,
\label{apbvocho}
\eeq
which, indeed,  is the same as in the one in eq. (\ref{ctvcinco}) for
this case. Notice that $n$, which in our present approach is  the number
of D-strings, corresponds to the quantization integer of the D3-brane
worldvolume gauge field. It follows that the minimal energy
configurations occur for
$\theta\,=\,\pi n/N$ and its energy density is the one written in eq. 
(\ref{ctvseis}).  This agreement shows that our ansatz represents 
D-strings growing up into a D3-brane configuration of the type studied
in the main text. 

Let us finally point out that the same ansatz of eqs.
(\ref{apbnueve})  and (\ref{apbtrece}) can be used to describe the
configurations in which D0-branes expand into a D2-brane in the NS5
background of eqs. (\ref{apbtres}) and (\ref{apbcuatro}). In this
case, which corresponds to the situation analyzed in ref. \cite{Bachas},
the D2-branes are located at fixed $r$ and one only has to compute the
determinant of the matrix (\ref{apbquince}) in the D0-brane action. By
using eq.  (\ref{apbvuno}) one easily finds the same hamiltonian and
minimal energy configurations as those of ref. \cite{Bachas}.

\end{document}